\documentclass{emulateapj}
\slugcomment{{\sc Accepted to The Astrophysical Journal:} October 28, 2010} 
\usepackage{amsmath}
\usepackage{xspace}
\usepackage{graphicx}
\usepackage{epstopdf}

\newcommand{\LyA}{Lyman $\alpha$\xspace}
\newcommand{\eps}[1]{\ensuremath{\epsilon = {#1}}\xspace}

\begin{document}
\title{Chemistry of a protoplanetary disk with grain settling and \LyA radiation}
\author{
Jeffrey K. J. Fogel\altaffilmark{1}, Thomas J. Bethell\altaffilmark{1}, Edwin A. Bergin\altaffilmark{1}, Nuria Calvet\altaffilmark{1}, Dmitry Semenov\altaffilmark{2}
}

\altaffiltext{1}{Department of Astronomy, University of Michigan, 830 Dennison Building, 500 Church Street, Ann Arbor, MI 48109; fogel@umich.edu, tbethell@umich.edu, ebergin@umich.edu, ncalvet@umich.edu}
\altaffiltext{2}{Max-Planck-Institute for Astronomy, Koenigstuhl 17, 69117 Heidelberg, Germany; semenov@mpia.de}

\begin{abstract}
We present results from a model of the chemical evolution of protoplanetary disks. In our models we directly calculate the changing propagation and penetration of a high energy radiation field with \LyA radiation included.  We also explore the effect on our models of including dust grain settling.  We find that, in agreement with earlier studies, the evolution of dust grains plays a large role in determining how deep the UV radiation penetrates into the disk.  Significant grain settling at the midplane leads to much smaller freeze-out regions and a correspondingly larger molecular layer, which leads to an increase in column density for molecular species such as CO, CN and SO.  The inclusion of \LyA radiation impacts the disk chemistry through specific species that have large photodissociation cross sections at 1216 \AA.  These include HCN, NH$_3$ and CH$_4$, for which the column densities are decreased by an order of magnitude or more due to the presence of \LyA radiation in the UV spectrum.  A few species, such as CO$_2$ and SO, are enhanced by the presence of \LyA radiation, but rarely by more than a factor of a few.  
\end{abstract}


\section{Introduction} \label{section-introduction}
Planets form in protoplanetary disks around low mass stars, though the exact mechanism of this formation is still not clear.  It is therefore important to understand the physical conditions in the disks in which they form.  In particular, studying the chemical history of these disks provides us with a wealth of information because the chemistry responds to the physical environment.  From the observed molecular emission we are able to explore disk physics such as the overall ionization fraction, which impacts on accretion physics, and disk kinematics, a direct probe of rotation and turbulence \citep[][and references therein]{semenov2010-review}.  In addition, the chemical composition of extrasolar protoplanetary disks can be compared to that of the many remnants from planet formation that we observe in our own solar system, such as comets, meteorites and asteroids.  The composition of these bodies provides a chemical memory of the formation of our solar system.  It is therefore necessary to understand the chemical evolution of protoplanetary disks so that we can unravel the mystery of how our own solar system formed and characterize planet formation elsewhere.  

Because chemical processes respond to the physical state of the gas, knowledge of the physical environment of the disk is required to properly model the chemistry.  This includes determining how the chemical composition of the disk is affected by the local density and temperature as well as by radiation from the central star and the surrounding environment.  Based on observations of T Tauri stars, a typical disk system consists of a central star surrounded by a flared dusty disk hundreds of AU in radius.  T Tauri stars are low mass young stellar objects, masses range from 0.08 - 2 M$_\sun$, with emission spectra that peak in optical wavelengths, though they also emit in the UV and X-ray regimes \citep{ardila2002, kastner1997}.  The optical emission determines the dust temperature within the structure of the disk \citep{calvet1991}, but it is the UV and X-ray radiation that provides the ionizing photons needed to power molecular chemistry.  Due to the reprocessing of stellar radiation, the disks themselves are flared \citep{kenyon1987}.  Armed with models of the disk physical structure, chemical models have suggested that beyond R $\sim$ 40 AU, the chemical structure of the disk can be divided into 3 regions: the photon dominated region, the warm molecular layer and the frozen-out midplane \citep[e.g.][]{aikawa2002, willacy2000, semenov2004}.  This is in agreement with the limits set by observations of molecular emission.  Radiation from the central star as well as the interstellar radiation field ionize atoms and dissociate molecules at the surface, leading to a photon dominated region at the top of the disk where chemical elements exist primarily as free atoms.  Near the midplane, the temperature is low enough that freeze-out of molecules as ices on grains can occur.  In between is a layer that is warm enough to prevent freeze-out, but shielded enough so that molecular species can form.  The exact locations of these regions depend on the chemical species being looked at, due to the differences in binding energies and freeze-out temperatures between different molecules.  

Over the years, chemical models of protoplanetary disks have grown increasingly more sophisticated \citep[][and references therein]{bergin2007-chemevo}.  The first generation determined the disk composition by assuming thermochemical equilibrium in the disk \citep[e.g.][]{lewis1974}.  Once it was realized that chemical kinetics, and not just the solids, played a large role in determining the observed abundances, models were developed that directly solved the time-dependent reaction equations \citep[e.g.][]{willacy1998, aikawa1999-2D}.  The current state-of-the-art models include cosmic ray and X-ray ionization, UV radiation from the central star, adsorption and desorption onto and off of grains, changing dust grain sizes, some simple mixing, surface chemistry and some isotopic chemistry \citep[e.g.][]{aikawa2008, willacy2009, woitke2009, nomura2009}.  

One major aspect of protoplanetary disk chemistry models is to properly treat the transfer of FUV radiation that is incident on the disk.  This includes UV radiation, both from external sources (typically the Interstellar radiation field, ISRF) and from the central star (mostly the accretion shock, which dominates the stellar radiation field, SRF, in the UV regime for T Tauri stars) as well as X-ray emission from the central star.  The two different sources of UV radiation have different angles of incidence on the disk with the ISRF having a normal angle of incidence, while the SRF has a very shallow angle of incidence \citep{willacy2000}.  Because of this, it has been shown that properly treating the absorption and scattering of the SRF is critical in calculating the disk chemical evolution as there is little penetration of the UV field into the disk without scattering being taken into account \citep{van-zadelhoff2003}.  

In addition to properly treating the UV radiation transfer, \LyA radiation is an important component in fully dealing with the UV field.  \citet{bergin2003, bergin2004} and \citet{herczeg2004} showed that \LyA radiation is dominant in the UV fields of some T Tauri stars, and in fact can carry up to 75 \% of the UV flux from these stars.  Additionally, the scattering of the \LyA radiation in the disk will be different from the scattering of the general UV field.  Whereas the general UV field scatters solely off of the dust grains, the \LyA radiation scatters from the atomic hydrogen in the disk prior to scattering off of dust grains, meaning that it needs to be treated separately from the rest of the UV field \citep{bethell2010}.  The scattering of \LyA radiation from atomic hydrogen also differs from the scattering off of dust grains in that it is isotropic, as opposed to scattering off of dust grains which is preferentially in the forward direction \citep{bonilha1979, draine2003}.  Previous models have included stellar UV fields, but no previous model has included an observationally motivated radiation field that coupled the resonant line radiation transfer of \LyA radiation with the full disk chemistry.  

Given the dominance of \LyA radiation in T Tauri stars, the lack of \LyA radiation in previous disk chemistry models is a significant omission \citep{bergin2003}.  In particular there are systematic differences in photodissociation rates for species that have cross sections at 1216 \rm{\AA} \citep{van-dishoeck2006}, as compared with those that do not, which will dramatically affect the chemistry of the disk.  For example, ratios such as CN/HCN will be affected since CN is dissociated primarily below 1150 \rm{\AA}  while HCN is subject to dissociation by \LyA radiation \citep{bergin2003}.  The inclusion of \LyA radiation should then increase the CN/HCN ratio and potentially allow for a better fit to the observed ratios \citep[e.g.][]{thi2004}.  

Another factor in properly dealing with UV radiation propagation is the dust in the disk.  Both dust settling and dust coagulation will lower the opacities in the upper regions of the disk, leading to a flatter overall disk and greater penetration of radiation \citep{dullemond2003}.  This increase in penetration depth for the UV radiation has been measured observationally by looking at the spectral energy distributions of these T Tauri stars and measuring the SED slope in the mid-IR wavelengths (5-30 $\mu$m), with more dust settling leading to a flatter slope \citep[e.g.][]{furlan2008}.  These observations show that most disks are fit by models with significant dust settling included.  

Motivated by these physical effects, we have created a protoplanetary disk chemistry model that includes the treatment of the radiation transfer of \LyA radiation from \citet{bethell2010} coupled with the full disk chemistry to demonstrate how the inclusion of \LyA radiation will affect disk chemistry.  We have also looked at the effect of dust settling on the chemical composition of the disk.  In \S \ref{section-model} we describe the model used to calculate the chemistry in these protoplanetary disks.  The physical model, as well as the types of chemical reactions used, is discussed.  In \S \ref{section-results} we present the effects that including dust setting and \LyA radiation has in terms of the chemistry of the disk and in \S \ref{section-discussion} we discuss why it is necessary to include them in future disk chemical models.

\section{Model} \label{section-model}
Our numerical model is based off of the ALCHEMIC code, written by \citet{semenov2010} for use on molecular cloud chemistry, and heavily modified for our purposes.  It is a 1+1D model, which means that while the code itself is one-dimensional (1D) over height at a specific radius, the code was run at a range of radii to give a pseudo-2D result.  The disadvantage of such a model is that there is no interaction between horizontal zones, though with no mixing included this should be a minor effect.  Our treatment of the UV field, however, was a full 2D treatment.  The code modeled a flared protoplanetary disk with incident radiation from a central T Tauri star as well interstellar cosmic rays.  Our model calculated the chemical abundances for 639 chemical species and 5910 chemical reactions.  While most of the reactions were not time-dependent, we did incorporate a few time-dependent reactions into the network which will be discussed in more detail below.  

\subsection{Physical Model}
\label{physical-model}
In this section we describe the physical structure of the disk that was used as well as the radiation field incident on the disk.  We used the parameters for a typical T Tauri star with radius $R_*$ = 2 $R_{\sun}$, mass $M_*$ = 0.5 $M_{\sun}$, mass accretion rate $\dot{M} = 10^{-8} M_{\sun} yr^{-1}$ and temperature $T_*$ = 4000 K \citep[e.g.][]{kenyon1995}.  The dust composition, taken from \citet{dalessio2006}, consists of two populations, a large grain and a small grain component.  In both cases, the grain size distribution is given by a power law of the grain radius (a): n(a) = n$_0 \left(\frac{a}{a_0}\right)^{-p}$, where p = 3.5, a$_{min}$ = 0.005 $\mu$m, a$_{max, small}$ = 0.25 $\mu$m and a$_{max, large}$ = 1.00 mm.  

\begin{figure*}
	\includegraphics[width=0.5 \textwidth, clip=true, trim=30 15 75 10]{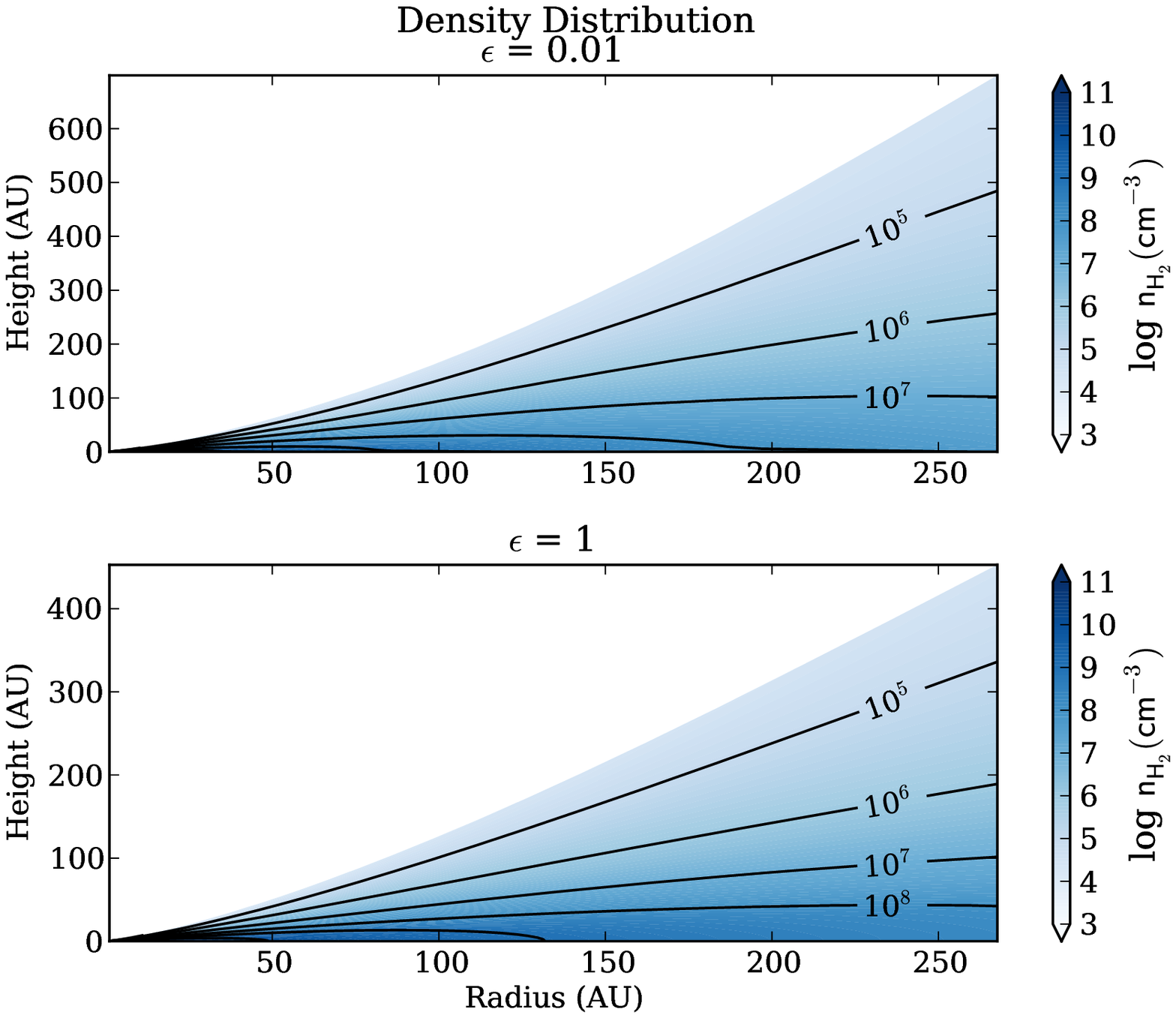}
	\includegraphics[width=0.5 \textwidth, clip=true, trim=25 15 80 10]{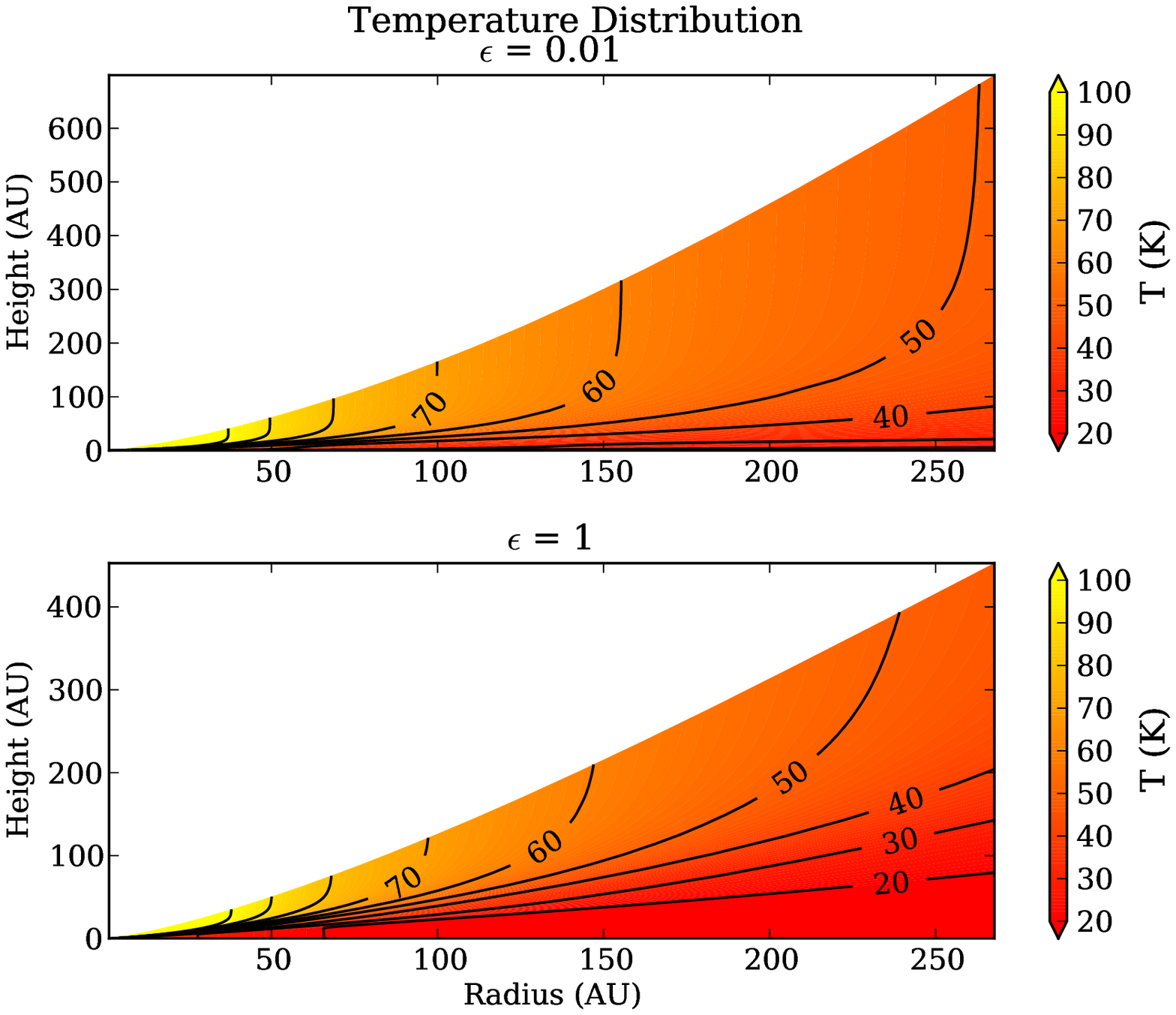}
	\caption{Density (\emph{left}) and Temperature (\emph{right}) structure of a disk model with \eps{0.01} and \eps{1}.  The model was run with the parameters for a typical T Tauri star with radius $R_*$ = 2 $R_{\sun}$, mass $M_*$ = 0.5 $M_{\sun}$, mass accretion $\dot{M} = 10^{-8} M_{\sun} \rm{yr^{-1}}$ and temperature $T_*$ = 4000 K \citep[e.g.][]{kenyon1995}.  The dust composition and settling were taken from \citet{dalessio2006} with a$_{max, small}$ = 0.25 $\mu$m and a$_{max, large}$ = 1.00 mm.}
	\label{structure}
\end{figure*}

Our model includes the effects of dust settling, parameterized by $\epsilon$, defined as the dust-to-gas mass ratio of the small grains in the upper regions of the disk relative to the standard dust-to-gas mass ratio \citep{dalessio2006}.  A value of \eps{1} means that no settling has taken place, while a smaller value of $\epsilon$ indicates a depletion of dust grains in the upper regions of the disk and a corresponding increase in large dust grains near the midplane.  We therefore assume that the disk is well shielded in a thin layer near the midplane.  The general disk temperature and density structure for dust settling parameters of \eps{1} and 0.01, shown in Figure \ref{structure}, are based on the assumption that the gas and dust temperatures are well coupled \citep{kamp2004}.  When dust settling is included, the temperature structure of the disk changes which leads to a less flared disk due to the surface density being approximated as 

\begin{equation}
\label{eqn-density}
\Sigma \sim \frac{\dot{M} \Omega_k}{3 \pi \alpha c_s^2(T)}
\end{equation}

\noindent
where $\Omega_k$ is the Keplerian angular velocity, $\alpha$ is a free parameter defined by \citet{shakura1973} and c$_s$ is the sound speed.  Because of the dependence of the sound speed on temperature, a different temperature structure will lead to a different density structure for the disk.  For the settled models, the temperature of the disk near the midplane is increased due to the greater penetration depth for radiation in the more settled disks, which will play a large role in determining the chemistry of the disk.  Recent studies have found that protoplanetary disk observations are well fit using dust settling parameters of $\epsilon$ $\lesssim$ 0.01 \citep{furlan2006}.  For the results presented in this paper, we compared $\epsilon$ values of 1, 0.1 and 0.01.  Based on the work done by \citet{andrews2005} we assume that the inner and outer disk dust populations are coupled, and so we treat them identically.  The mass of the disk depends on the dust settling parameter and was 0.02 M$_\Sun$ for \eps{1} and 0.01 M$_\Sun$ for \eps{0.1} and 0.01.

In order to more clearly understand the gross effects of \LyA radiation on the chemistry of the disk we do not include any turbulent mixing in the disk.  Additionally, grain growth in our model is not time dependent, but is fixed for each model.  While including these time-dependent processes in a detailed chemical model might be interesting, even with modern computers it is a very computationally intensive set of calculations.  Many previous models of disk chemistry have reproduced results in agreement with observations without including these processes \citep[e.g.][]{aikawa1999-2D, willacy2000, aikawa2006}.  In comparison, to first order diffusion smoothes out abundance variations with height and increases the depth of the molecular layer \citep[e.g.][]{ilgner2004, semenov2006, willacy2006}.  Thus we can still understand the large-scale effects of \LyA radiation on the chemistry without including diffusion or turbulent mixing in our model.

\subsection{UV Field Calculation}
\label{sec-uvfield}
\begin{figure}
	\includegraphics[width=0.5 \textwidth, clip=true, trim=25 5 30 30]{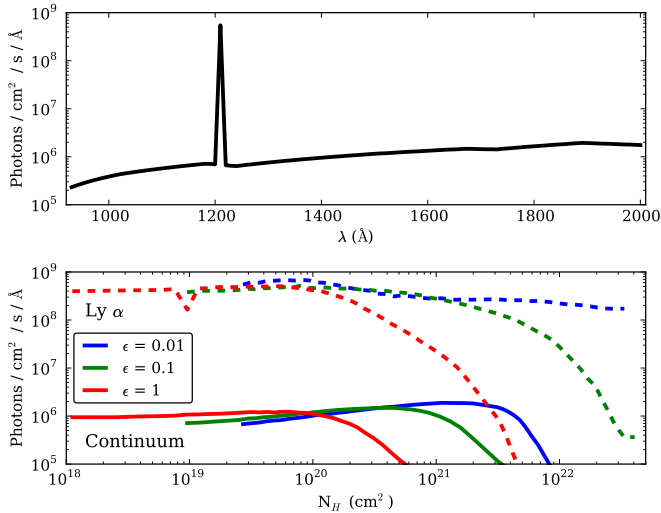}
	\caption{{\em Top Panel - } Sample stellar UV field from \citet{bethell2010}.  This example was taken from just below the top of the disk for a radius of 208 AU and an $\epsilon$ of 0.01.  
	{\em Bottom Panel - } The effects of scattering and attenuation on the UV field in the disk.  Solid lines are the continuum, dashed lines are \LyA radiation.  }
	\label{uvfield}
\end{figure}

The UV field for the central star that was used for our model was calculated by \citet{bethell2010}, who explored the radiative transfer and scattering of \LyA radiation in a protoplanetary disk.  The UV field was calculated such that the G$_0$(100 AU) value, the scaling factor between the total flux of the ISRF in the FUV band and the total flux in our radiation field at 100 AU \citep{habing1968}, was set to $\sim$ 700.  This value was chosen as an average of the G$_0$ values observed by \citet{bergin2004} for a set of T Tauri stars.  An example of the UV spectra used can be seen in the top panel of Figure \ref{uvfield}, which shows the UV field for a radius of 208 AU for an \eps{0.01}.  The UV field was assumed to be very smooth for all wavelengths other than \LyA, which completely dominates the spectra.  While there are known emission lines in the other portions of the UV spectra \citep{bergin2003}, they are orders of magnitude lower than the \LyA radiation and would not significantly modify the disk chemistry presented here.  The lower panel of Figure \ref{uvfield} shows the dependence of the UV field on the column density of the disk.  Due to geometrical dilution, the UV field strength actually peaks slightly inside the disk and not at the surface.  The distance from the star to a point in the disk is r$^2$ + z$^2$.  As z decreases, this distance shrinks and the UV field strength increases until there is enough material between the star and the point for the UV field to be attenuated.  

The \LyA photon density was calculated in the two-stage process described in \citet{bethell2010}.  Here we provide only a high-level description of the aspects of the method most relevant to disk chemistry. In order to follow continuum FUV and \LyA photons from the star into the disk, the gas and dust opacity must first be known.  We assume that the pure absorption of both continuum and \LyA photons is simply due to dust.  The scattering opacity requires more careful consideration and is of central importance to the propagation of \LyA.  For continuum photons we assume scattering is due to dust.  For \LyA the scattering is a combination of dust scattering and resonant scattering by H atoms.  For consistency, we use the same dust grain population as that used in \citet{dalessio2006}.  The first calculation is therefore to establish the distribution of atomic H, which is done by balancing the formation of H$_2$ on grains with its destruction by UV photons in the Lyman-Werner bands (912-1100\AA, \citealt{black1987}).  Following \citet{spaans1997} we write the steady-state H$_2$ fraction,  $f_{\textrm{H}_2}$, as

\begin{equation}
\label{tom_eqn}
\frac{n(\textrm{H}_2)}{n_H}=\frac{Rn_H}{\zeta+2Rn_H}.
\end{equation}

We adopt the \citet{cazaux2004} expression for the grain-surface H$_2$ formation rate R.  The simple `shielding function' approach of \citet{draine1996} is used to estimate the H$_2$ dissociation rate, $\zeta$, ensuring consistency with later chemical computations (\S \ref{photodissociation}).  The photons are propagated through the (irregular 2D) spatial grid using standard Monte Carlo radiative transfer techniques \citep[e.g.][]{bethell2007}.  This approach allows the potentially important effects of scattering to be included \citep[e.g.][]{van-zadelhoff2003}.  The radiative transfer through the dusty H/H$_2$ distribution is iterated until a converged, simultaneous solution is achieved for both the H/H$_2$ distribution and UV radiation field.  This iterative approach is typical for problems in which the opacity is coupled to the radiation field. It is important to note that the H/H$_2$ transition occurs readily, well above the regions where the bulk of the heavy element chemistry takes place.  The H/H$_2$ transition defines the top of the so-called 'warm molecular layer'.  In this sense the chemistry occurs against a background of a fully molecular H$_2$ disk.  However, before the \LyA photons can reach these depths they must first pass through the upper `photodissociation layer', composed of mostly atomic H.  This leads us to the second stage of the calculation; the detailed propagation of \LyA photons.  We are in essence repeating the UV transport calculation, but this time treating the \LyA propagation in greater detail.   While this two-stage approach is not fully self-consistent, it renders the overall problem more tractable.

\begin{figure}
	\includegraphics[width=0.5 \textwidth]{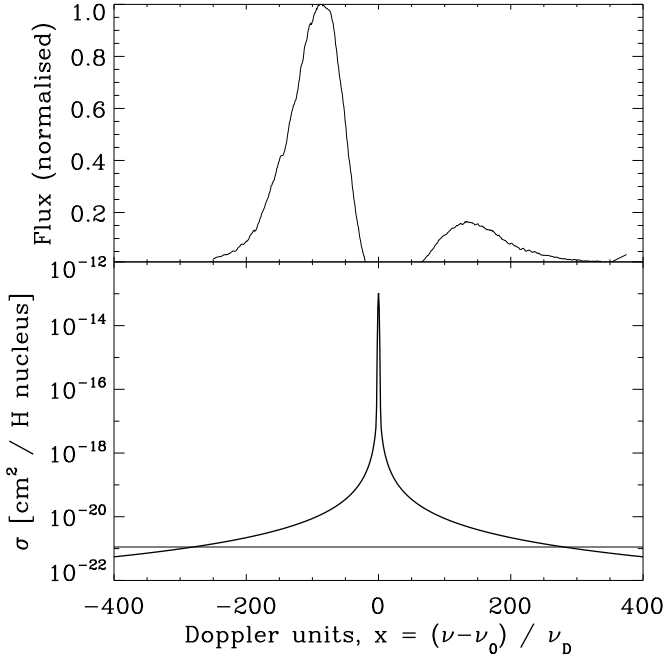}
	\caption{The stellar \LyA line profile of TW Hya (upper panel, taken from \citealt{herczeg2002}), and resonant scattering (Voigt) profile due to HI atoms in the disk (lower panel).  The horizontal line in the lower panel represents the dust opacity.}
	\label{lya_profile}
\end{figure}

Although the shape of the stellar \LyA line bathing the disk in T Tauri systems is generally not observed directly, it is known to be very wide in the case of TW Hya and in other stars with molecule-rich disks (FWHM $>$500km s$^{-1}$, \citealt{herczeg2002, bergin2004}), especially when measured in Doppler widths of the disk gas (Figure \ref{lya_profile}).  As a result, the vast majority of \LyA photons scatter off the wings of the resonant-scattering Voigt profile presented by HI in the photodissociation layer.  Even at frequency offsets as large as $\pm$200 Doppler widths relative to line center (1216\AA), the resonant scattering opacity of H atoms suspended above the disk is greater than that due to the accompanying dust (assuming the typical interstellar dust:gas mass ratio $\sim0.01$, i.e. $\epsilon=1$).  Of course, this difference is further enhanced by the removal of dust via settling. Dust settling also enhances the abundance of atomic hydrogen by reducing the H$_2$ formation rate (Eqn. \ref{tom_eqn}).  From the point of view of a stellar \LyA photon it is as though the disk is covered in an optically thick layer of high-albedo material (the atomic hydrogen), which immediately scatters approximately half of the incident \LyA photons out into space and the remaining photons downward into the disk.  This downward scattering greatly increases the penetrating power of \LyA photons compared to the feebly dust-scattered continuum photons.  Due to the great width of the \LyA line, its overall frequency evolution as the photons diffuse into the disk is relatively unimportant, and the scattering process can be treated as conserving photon energy.  Nevertheless, in our calculation we follow the frequency evolution of photons using the detailed theory of  `partial frequency redistribution' \citep{hummer1962}.  Ultimately these photons are lumped together into a bolometric \LyA photon density, J$_{\lambda}$(r,z), that is used later in chemical rate equations.   

Because of the proximity of TW Hya, and the lack of interstellar gas along the line of site,  it is the only system with detectable Ly $\alpha$ beyond the broad wings.  However, H$_2$ fluourescent emission is clearly detected in the handful of low mass T Tauri disk systems with high resolution FUV spectra.   This H$_2$ is pumped via coincidences with \LyA \citep{herczeg2002}.  Thus it is likely that \LyA emission is an important facet of the UV field in {\em most if not all} accreting systems, although the exact strength amongst systems needs to be determined.  This assertion will be quantified with new UV spectra from the {\em Cosmic Origins Spectrograph} on HST, but preliminary results are in agreement \citep{herczeg-privcom}.

In order to determine the effects of \LyA radiation on the chemistry of the disk, we ran two models, one with and one without the \LyA radiation.  In the case without \LyA radiation the UV field was identical, except that the continuum was interpolated over the \LyA peak to remove it.  The results from these models will be presented in \S \ref{section-results}.  

\subsection{Chemical Model}
For our chemical model we found the abundance of each species in the disk by solving rate equations for each chemical species at each time-step and zone in the disk.  The basic equation we solved can be written as

\begin{equation}
\frac{dn(i)}{dt} = \sum_j\sum_l k_{j l} n(j) n(l) - n(i)\sum_j k_{i j} n(j)
\end{equation}

\noindent
where $n(i)$ is the abundance of species $i$ and $k_{i j}$ is the reaction rate for reacting species $i$ and $j$.  At its most basic, this equation is a summation of the rates of destruction of a species subtracted from the summation of the rates of formation.  The specific rate equations for each reaction used in our network depended on the type of the reaction and will be discussed in further detail below.  Three-body reactions were not included in our network, as we were mostly concerned with the outer regions of the disk and three-body reactions become competitive only in the inner regions of the disk where temperatures and densities are very high.  Initial abundances of each species were taken to be molecular cloud abundances, approximated from \citet{aikawa1999-2D} and are listed in Table \ref{initabun}.  

\begin{deluxetable}{lrlr}
\tablewidth{0pt}
\tablecaption{Initial Abundances\tablenotemark{a}\label{initabun}}
\tablehead{
\colhead{Species} & \colhead{Abundance} & \colhead{Species} & \colhead{Abundance}
}
\startdata
H & 1.00 		& CO  & 1.00E-4 \\
He & 1.40E-1 	& N$_2$ & 1.00E-6 \\
N & 2.25E-5 	& C & 7.00E-7 \\
CN & 6.00E-8 	& NH$_3$ & 8.00E-8 \\

H$_3$$^+$ & 1.00E-8 	& HCN & 2.00E-8 \\
S$^+$ & 1.60E-6 		& C$^+$ & 1.00E-8 \\
Si$^+$ & 1.60E-9 		& HCO$^+$ & 9.00E-9 \\
Mg$^+$ & 3.00E-8 		& H$_2$CO & 8.00E-9 \\
Fe$^+$ & 2.00E-8 		& C$_2$H & 8.00E-9 \\ 

H$_2$O(gr) & 2.50E-4	& CS & 2.00E-9 \\
GRAIN & 6.00E-12		& 

\enddata
\tablenotetext{a}{The initial abundances are by number density relative to the total Hydrogen abundance}
\end{deluxetable}

Our chemical network was based on the Ohio State University Astrophysical Chemistry Group gas-phase model from March 2008\footnote{http://www.physics.ohio-state.edu/$\sim$eric/research.html} \citep{smith2004},  with a number of additions that will be elaborated on below.  This network includes some gas-grain interaction, electron-grain recombination, cosmic ray ionization, cosmic-ray induced photoreactions, ion-molecule reactions, charge exchange reactions, negative ion - neutral species reactions, radiative association, associate ejection, neutral + neutral $\rightarrow$ ion + electron, neutral-neutral chemical reactions, neutral-neutral radiative association, dissociative recombination, radiative recombination, positive ion-negative ion recombination, electron attachment, photoionization and photodissociation.  The reaction rates for all of these reactions can be found on the OSU website, though only about 10-20\% of these reaction rates have been measured in the laboratory, which adds a certain amount of uncertainty to all chemical network calculations \citep[e.g.][]{wakelam2006, vasyunin2008}.  

\subsubsection{Photodissociation}
\label{photodissociation}
One addition to the OSU network was the incorporation of photodissociation of molecules from measured cross sections in the literature.  These rates are dependent on both the strength and the shape of the radiation field that is used for the model.  Because we were using an observationally motivated radiation field, it was necessary to properly treat the molecular photodissociation.  There were two different methods that were used to calculate the photodissociation cross sections.  For those molecules where UV cross sections for the molecules have been measured \citep{van-dishoeck2006}, the reaction rate was determined by

\begin{equation}
\label{ourprate}
k_{photodissociation} = \int \frac{4 \pi \lambda}{h c} \sigma(\lambda) J_\lambda(r,z) d\lambda  \; s^{-1}
\end{equation}

\noindent
where $\sigma(\lambda)$ is the wavelength-dependent cross section of the molecule and $J_\lambda(r,z)$ is the radiation field at any point in the disk (calculation describe in \S \ref{sec-uvfield}).  For all of the species that did not have measured cross sections in the literature, the photodissociation rates were calculated by

\begin{align}
\label{simpprate}
& k_{photodissociation} = \nonumber \\
& G_0(100 AU) \left(\frac{100.0}{R(AU)}\right)^2 \alpha e_{\tau}(r,z,\lambda=1500\rm{\AA})  \; s^{-1}
\end{align}

\begin{equation}
e_{\tau} = \frac{J_{\lambda,z}}{J_{\lambda, z_{max}}}	\label{eqn-etau}
\end{equation}

\noindent
$G_0(100 \rm{AU})$, as defined above, is the scaling factor between the ISRF and the radiation field we used for our model and was set to $\sim$ 700.  $\alpha$ is the unshielded photodissociation rate taken from the UMIST 2006 database \citep{woodall2007}.  $e_{\tau}$ accounts for the attenuation of the radiation field.  Normally, this term would just be e$^{-\tau}$, however due to scattering of the radiation the attenuation needed to be calculated as the radiation field for a given height z, ($J_{\lambda,z}$), divided by the maximum radiation field at that radius ($J_{\lambda, z_{max}}$).  1500 \rm{\AA} was chosen as a wavelength well beyond that of \LyA to use for this calculation.  

\begin{figure}
	\includegraphics[width=0.5 \textwidth, clip=true, trim=10 10 25 20]{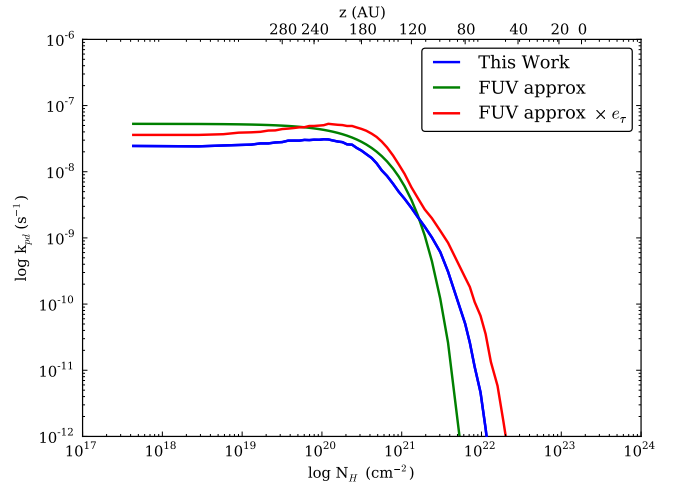}
	\caption{Plot of three different photodissociation rates of HCN at radius R = 208 AU and \eps{1}.  The dashed line (FUV approx) assumes a normal ISRF scaled to match the FUV flux of a typical T Tauri star without any scattering included.  The dotted line (FUV approx $\times$ e$_{\tau}$) includes e$_{\tau}$, which incorporates the attenuation and scattering of the UV field in the disk.  The solid line is the result of calculating the photodissociation rate from the wavelength-dependent cross section of HCN.  }
	\label{pratecomp}
\end{figure}

A comparison of the calculated photodissociation rates of HCN vs. total column is shown in Figure \ref{pratecomp}.  This figure illustrates the reaction rate of HCN for three different methods of calculating the photodissociation rate.  The dashed line shows the result if we assume a normal ISRF scaled to match our radiation field in the FUV, Equation (\ref{simpprate}), but without any scattering included.  The dotted line includes the e$_{\tau}$ term which incorporates the attenuation and scattering of our UV field at 1500 \rm{\AA} \citep{van-zadelhoff2003, bergin2003}.  Finally, the solid line is the result of using Equation \ref{ourprate} to calculate the rate from the wavelength-dependent cross section and our UV field.  The spread in the three lines at high column densities indicates the importance of including scattering of the UV field in these calculations.  The difference between the solid and dotted lines occurs as a result of using the wavelength dependent cross section instead of just approximating it at 1500 \rm{\AA}, which is especially important for species with a cross section at 1216 \rm{\AA} that will be affected by the presence of \LyA radiation. Because of geometrical dilution due to the flaring of the disk, the maximum UV field is actually a few tens of AU below the surface.  This effect causes the rate calculated from the ISRF alone to be slightly higher in the lower column density regions of the disk.  

\subsubsection{Self-shielding of CO and H$_2$}
Because the photodissociation rates of CO and H$_2$ are dominated by line rather than continuous absorption, self-shielding of these two molecules becomes significant in the regions where they are optically thick.  For our model, CO self shielding was calculated from the rates given in \citet{visser2009-CO} while the H$_2$ self shielding rates were calculated from \citet{lee1996}.  The H$_2$ photodissociation rate was 

\begin{align}
k_{H_2 SS} =&\ 4 \times 10^{-11} G_0(100 AU) \left(\frac{100.0}{R(AU)}\right)^2 \nonumber \\
 &\times f \times e_{\tau}(r,z,\lambda=1000\rm{\AA})  (s^{-1})
\end{align}

\noindent
with f = $\left(\frac{N_{H_2}}{10^{14} cm^{-2}}\right)^{-0.75}$ for $N_{H_2} > 10^{14} \: cm^{-2}$ and f = 1 otherwise.  e$_{\tau}$ is defined in Eqn. \ref{eqn-etau}, though in this case it is calculated at 1000 \rm{\AA} since that is the middle of the range where H$_2$ is photodissociated \citep{van-dishoeck1988}.  The $G_0(100 AU)$ factor is used since the equation assumes an ISRF, which needed to be scaled up to match the UV field in our model.  

For CO, the shielding functions in \citet{visser2009-CO} were calculated for the unmodified ISRF and thus needed to be scaled up to match our UV field.  However, there was one modification to the scaling technique used to calculate the self-shielding rate for H$_2$.  Because CO does not have a cross section coincident with \LyA radiation at 1216 \rm{\AA}, but the radiation field used for our models had a large fraction of its UV flux at 1216 \rm{\AA}, using a G$_0$ value calculated with \LyA included would lead to erroneous results.  Instead, the peak at 1216 \rm{\AA} was removed by linearly interpolating over that portion of the curve and that resulting curve was used to calculate a new G$_0$ value used in the CO photodissociation calculations.  This led to a G$_0$(CO) value of $\sim$150.  In addition, e$_{\tau}$(r,z,$\lambda=960\rm{\AA}$) was used, with 960 \rm{\AA} chosen to be in the middle of the CO photodissociation range.  In both of the self-shielding cases the calculation of the rates were time-dependent since they depended on the column density of the molecule.  

\subsubsection{X-ray ionization and Cosmic-Ray ionization}
Observations of T Tauri stars have shown that they emit strongly in X-rays, with luminosities between 10$^{29}$ - 10$^{31}$ erg s$^{-1}$ \citep{flaccomio2009, feigelson1999, glassgold1997}.  \citet{aikawa1999-2D,aikawa2001} discussed the effects that this radiation has on protoplanetary disk ionization and chemistry, which includes secondary ionization of hydrogen gas as well as UV photolysis induced by the X-rays.  For our model, both of these effects have been included using the formalisms developed by \citet{glassgold1997} and \citet{aikawa2001} and using an X-ray luminosity of 10$^{29}$ erg s$^{-1}$.  For the cosmic ray ionization reactions we have followed the formalism presented by \citet{semenov2004} with cosmic rays striking the disk vertically on both sides of the disk with an attenuation column of 96 g cm$^{-2}$ \citep{umebayashi1981} and an unshielded cosmic-ray ionization rate of 1.3 $\times$ 10$^{-17}$ s$^{-1}$.  

\subsubsection{Adsorption \& Desorption}
The majority of our chemical network is made up of gas species and most of the reactions in our network are between two such species.  Just looking at gas species, however, leaves out a critical component of disk chemistry, reactions between gas species and dust grains.  In order to self-consistently treat the gas-grain interactions, the reaction rates for all such reactions was decreased by a factor of $\epsilon$ in order to take into account the effect of the dust grains settling to the midplane.  

Adsorption is the process by which a gas-phase species sticks to the surface of a dust grain and is retained due to van der Waals or surface bonding forces.  For the low temperatures that we were looking at in our models we assumed sticking coefficients of unity \citep{burke1983, bisschop2006}.  The reaction rate for species adsorption onto grains can then be written as

\begin{equation}
\label{eqn-adsorption}
k_{ad} = \sigma_{gr} \sqrt{\frac{8.0 k_B T}{\pi \beta m_H}} S n_{gr}  (s^{-1})
\end{equation}

\noindent
where $\sigma_{gr}$ is the cross section of the dust grain, approximated to be $\pi r_{gr}^2$.  The dust grains were assumed to have a radius of $r_{gr} = 0.1 \mu m$ which leads to a cross section of $3.14 \times 10^{-10} cm^{2}$ in Equation (\ref{eqn-adsorption}).  This is slightly larger than the average dust grain size calculated from the distribution in \S \ref{physical-model} of 0.02 $\mu$m due to the fact that smaller grains are subject to thermal spikes and it is therefore harder for molecules to adsorb onto small grains \citep{leger1985}.  $k_B$ is Boltzmann's constant, $\beta$ is the molecular weight of the species, and $S$ is the sticking coefficient. 

In our model there were three different methods of desorption, or ways to overcome the binding energy that holds a species to the surface of a dust grain after it has been adsorbed.  

\noindent
{\em Thermal Desorption.}  For thermal desorption, the rate of desorption was calculated using the Polyani-Wigner relation:

\begin{equation}
k_{td} = \sqrt{\frac{ 3.0 \times 10^{15} k_B E_b}{\pi^2 \mu m_H}} e^{-E_b/T} (s^{-1})
\end{equation}

\noindent
where $E_b$ is the binding energy of the molecule and $T$ is the dust temperature.  Binding energies were taken from \citet{hasegawa1993} and \citet{willacy2007} with the exception of H which was set to 1500 K, consistent with \citep{cazaux2004}.  By equating the flux of thermally desorbing molecules from a grain surface to the flux of adsorbing molecules, it is possible to estimate the freeze-out temperature for a given species.  Doing this, \citet{hollenbach2009} estimated freeze-out temperatures for H$_2$O of $\sim$100 K, but only $\sim$20 K for CO.  This has two implications for protoplanetary disk chemistry.  The first is that each species will have a slightly different disk structure, in terms of where we see the freeze-out region, that is dependent on the freeze-out temperature of that species \citep{aikawa2002}.  Secondly, the different temperatures mean that species will freeze-out at different locations in the disk, affecting which species are available for further chemical reactions in those regions.  

{\em Cosmic-ray Desorption.}  We included cosmic-ray desorption of ices from dust grains, however this is not a significant desorption mechanism in all regions of the disk.  As mentioned before, we assumed an attenuation column of 96 g cm$^{-2}$ for cosmic rays.  This means that at radii less than $\sim$ 1 AU, where $\Sigma$ is greater than 96 g cm$^{-2}$ even at the midplane, cosmic ray desorption never plays a large role in desorption.  In contrast, in the outer regions of the disk the cosmic ray attenuation is fairly low and that, combined with low temperatures at the midplane, means that there are regions of the disk where cosmic ray desorption is an important desorption mechanism.  For our model, the cosmic-ray desorption rate was calculated using the formalism of \citet{hasegawa1993} and \citet{bringa2004}.  

{\em Photodesorption.}  Photodesorption is the dominant form of desorption in the regions of the disk where the UV field can reach and the temperature is below a species' sublimation temperature.  The rate for photodesorption by UV photons was calculated as:

\begin{equation}
\label{PDrate}
k_{photodesorption}  = F_{UV} Y \frac{\sigma_{gr}}{N_{sites}} N_p \frac{n(i)}{n_{ice}} N_m^{-1}
\end{equation}

\noindent
where $F_{UV}$ is the total UV flux, $Y$ is the yield (the number of adsorbed particles ejected per incident photon), $\sigma_{gr}$ is the grain cross section, $N_{sites}$ is the number of reaction sites on the grain, assumed to be 10$^6$, $N_p$=2 is a correction for the fact that UV photons only penetrate the first few monolayers, $n(i)$ is the abundance of a given species, $n_{ice}$ is the total abundance of all species frozen onto grain surfaces and $N_m$ is the number of monolayers.  This equation is valid for an assumed two or fewer monolayers per grain, as laboratory experiments by \citet{oberg2007} indicate that most photodesorption occurs in the upper $\sim$ 2 monolayers.  All species were assumed to have a yield of $10^{-3}$, except for CO and H$_2$O for which we used  recently measured values from \citet{oberg2007, oberg2009-H2O}.  

In order to properly treat the \LyA radiation, the normal photodesorption reactions mentioned above were calculated with the \LyA radiation removed from the UV field used in $F_{UV}$ and instead a separate set of photodesorption reactions were included for all species that had a photodissociation cross section at \LyA wavelengths.  There is little experimental data available on photodesorption rates, so we assumed that the physics of species on dust grains is similar to the physics of those species in the gas phase.  This means that any species with a photodissociation cross section at 1216 \rm{\AA} was assumed to have a related \LyA photodesorption cross section.  In order to estimate the photodesorption yields of these species, we assumed a constant normalization factor between the gas-phase photodissociation cross sections in \citet{van-dishoeck2006} and the photodesorption yields.  To determine the normalization factor we matched the photodissociation cross section for H$_2$O of $1.2 \times 10^{-17} cm^{2}$ at $\lambda$ = 1216 \rm{\AA} with the photodesorption yield for H$_2$O(gr) of $\rm{2.36 \times 10^{-3}}$ measured by \citet{oberg2009-H2O}.  In symbolic form:

\begin{equation}
\frac{Y_{H_2O}}{Y_X} = \frac{\sigma_{H_2O}}{\sigma_X}
\end{equation}

\noindent
where $\sigma_X$ is the photodissociation cross section of species X and $Y_X$ is the photodesorption yield of that same species.  This new yield was then used in Equation (\ref{PDrate}), but this time with only the \LyA radiation flux included.  By treating the \LyA photodesorption separately in this manner, we remove the problem of the UV field having most of the flux at \LyA wavelengths, even if the species we are looking at does not have a cross section at 1216 \rm{\AA}.  

\subsubsection{Grain Surface Chemistry}
While the vast majority of our reactions involved gas species or molecules adsorbing onto or desorbing off of grains, there were two series of grain surface chemistry reactions that were included in the model.  The first is the formation of H$_2$ on grain surfaces.  Here we show the reaction and the formalism for computing the evolution:

\begin{multline}
	H + grain \rightarrow H(gr) + grain \\
	\frac{dn_{H(gr)}}{dt} = n_H n_{gr} <\sigma v> S + ...
\end{multline}
\begin{multline}
	\label{eqn-hgr}
	H(gr) + H 	\rightarrow H_2 \\
	\shoveleft{\text{if } n(H(gr)) > n(gr):} \\
	\frac{dn_{H_2}}{dt} = n_{gr} n_H <\sigma v> S + ...  
\end{multline}
\begin{multline}
	\text{if } n(gr) > n(H(gr)): \\
	\frac{dn_{H_2}}{dt} = n_{H(gr)} n_H <\sigma v> S + ... 
\end{multline}

\noindent
The reaction is calculated in two steps because the limiting step is for a hydrogen atom to find a grain.  Additionally, if the abundance of hydrogen atoms on grains, H(gr), is greater than the abundance of grains in the system (Equation (\ref{eqn-hgr})), the reaction rate is reduced by the ratio $\frac{n(gr)}{n(H(gr))}$ to take into account the fact that a gas phase atom must collide with a grain in order for the reaction to take place.  In all cases, the rate coefficient is $<\sigma v_H> S_H$ where $\sigma$ is the collisional cross section, defined as above as $\pi r_{gr}^2$, $v_H$ is the velocity of a hydrogen atom assuming a Maxwellian distribution at the gas temperature, and $S_H$ is the sticking probability of a hydrogen atom, assumed to be one.  

The second series of grain surface reactions was oxygen chemistry to form water on grains and followed the treatment from \citet{hollenbach2009}.  This included:

\begin{align}
O(gr) + H 		&\rightarrow OH(gr) \\
H(gr) + O 		&\rightarrow OH(gr) \\
H(gr) + OH 	&\rightarrow H_2O(gr) \\
OH(gr) + H 	&\rightarrow H_2O(gr)
\end{align}

\noindent
Laboratory experiments suggest that there is rapid formation of water ice on grain surfaces \citep{miyauchi2008}, which makes these reactions a necessary addition to the chemical network since water is one of the dominant oxygen-bearing species.  No other reactions on grains were included.

\section{Results} \label{section-results}
\subsection{Basic Model}
For the purposes of the discussion to follow, X(gr) indicates species X adsorbed onto a grain surface.  All abundances are relative to the total abundance of Hydrogen (n(H) + 2 n(H$_2$)) by number density.  In addition, all of the contour plots have been mirrored over the midplane to more easily display the features of the disk.  Unless otherwise stated, all values are taken at a time of 10$^6$ years.

\begin{figure}
	\includegraphics[width=0.5 \textwidth, clip=true, trim=20 10 30 20]{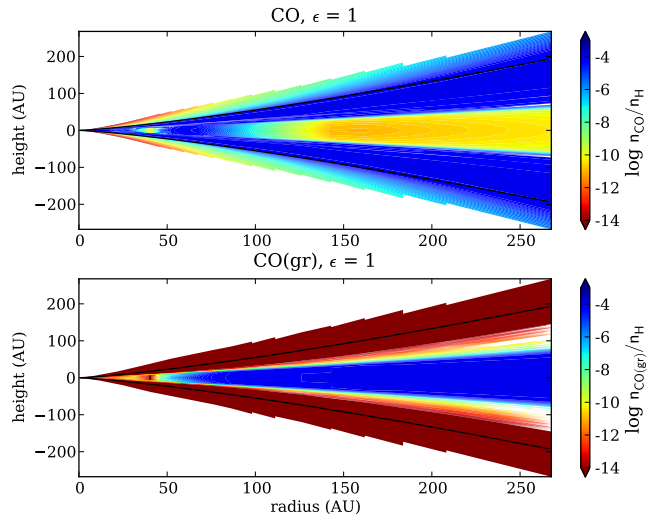}
	\caption{Contour plot of CO and CO(gr) abundances for a model with no dust settling included.  The dashed lines indicate the $\tau$ = 1 surface in the disk.  }
	\label{cotwoplot}
\end{figure}

\begin{figure*}
	\includegraphics[width=0.5 \textwidth ]{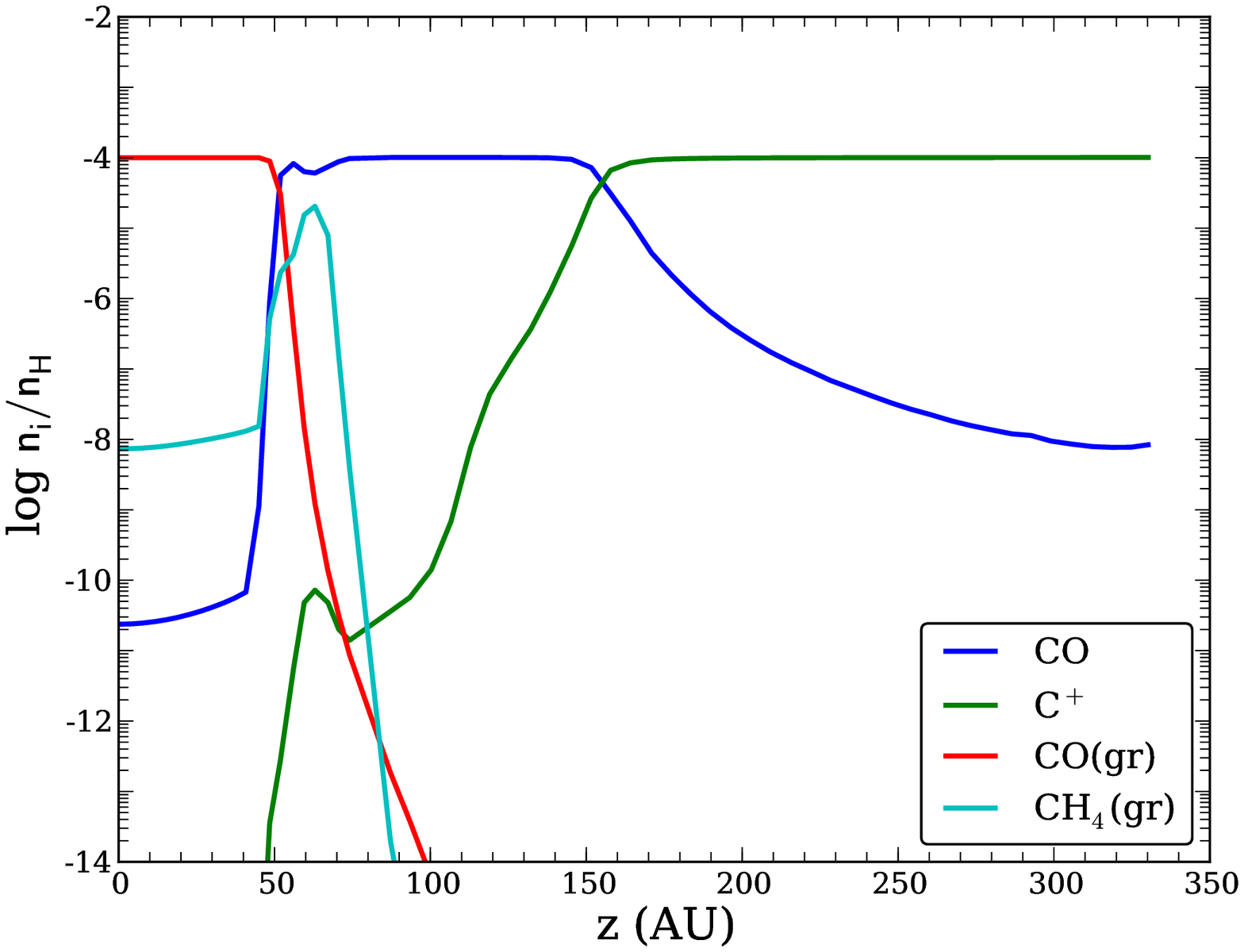}
	\includegraphics[width=0.5 \textwidth ]{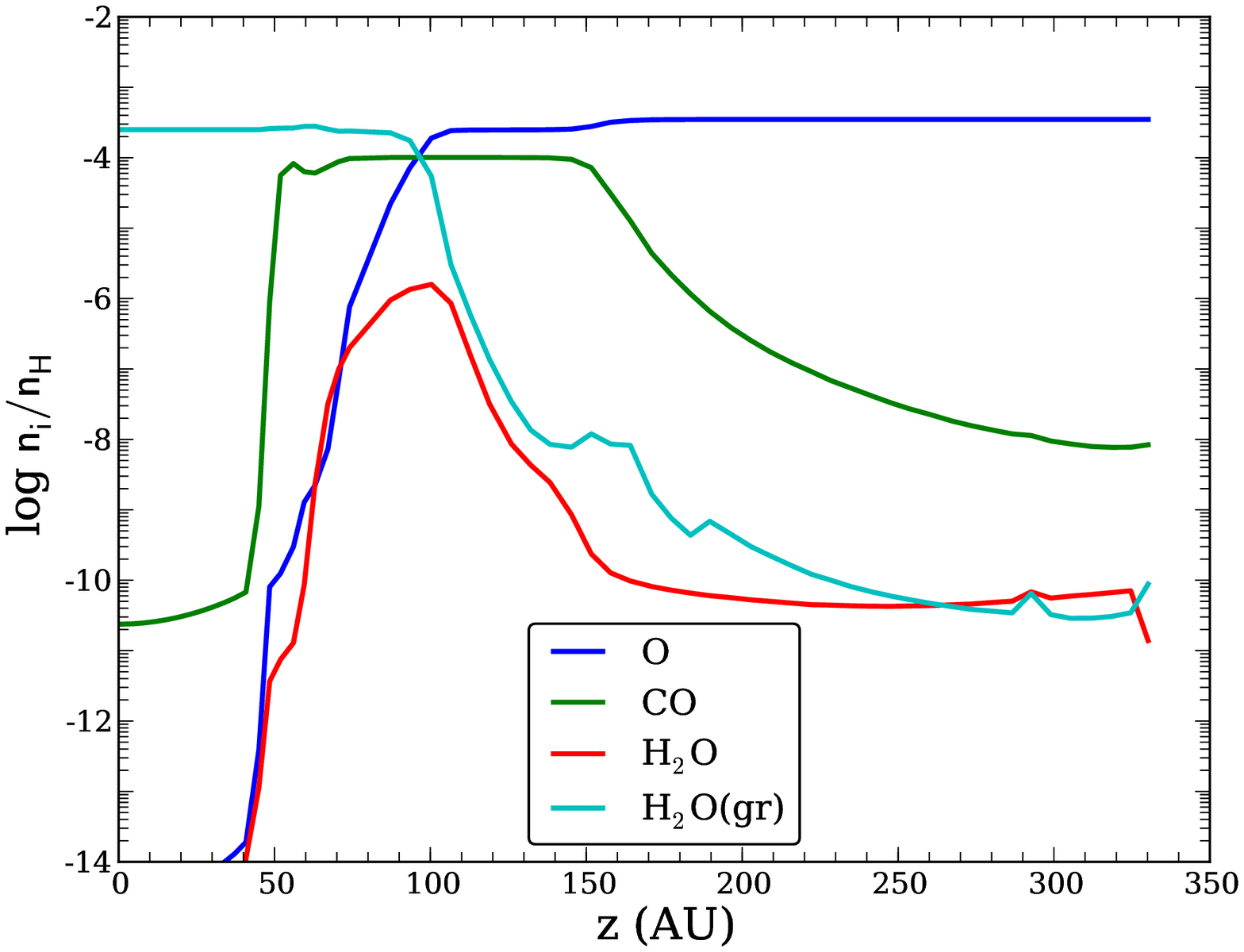}
	\includegraphics[width=0.5 \textwidth ]{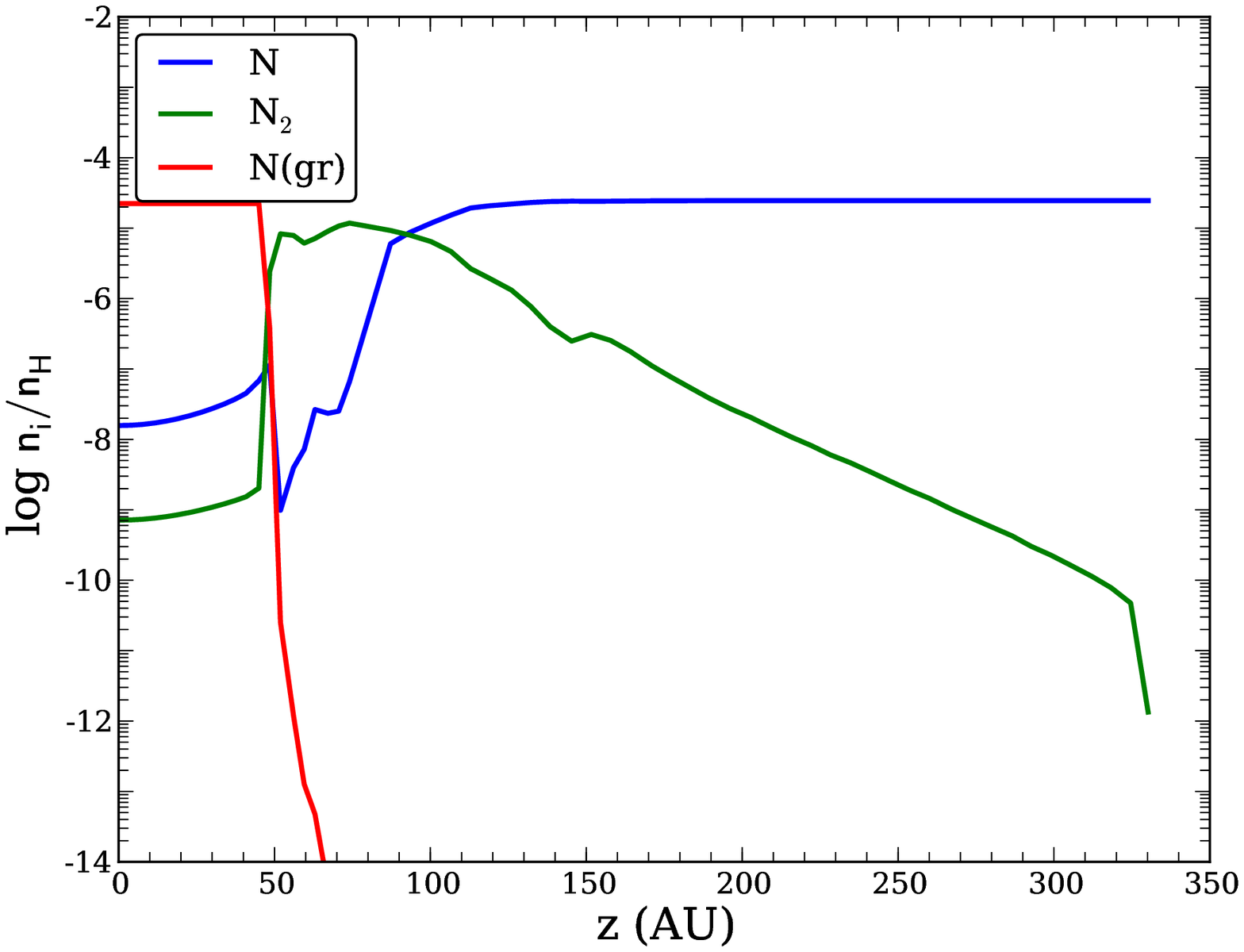}
	\includegraphics[width=0.5 \textwidth ]{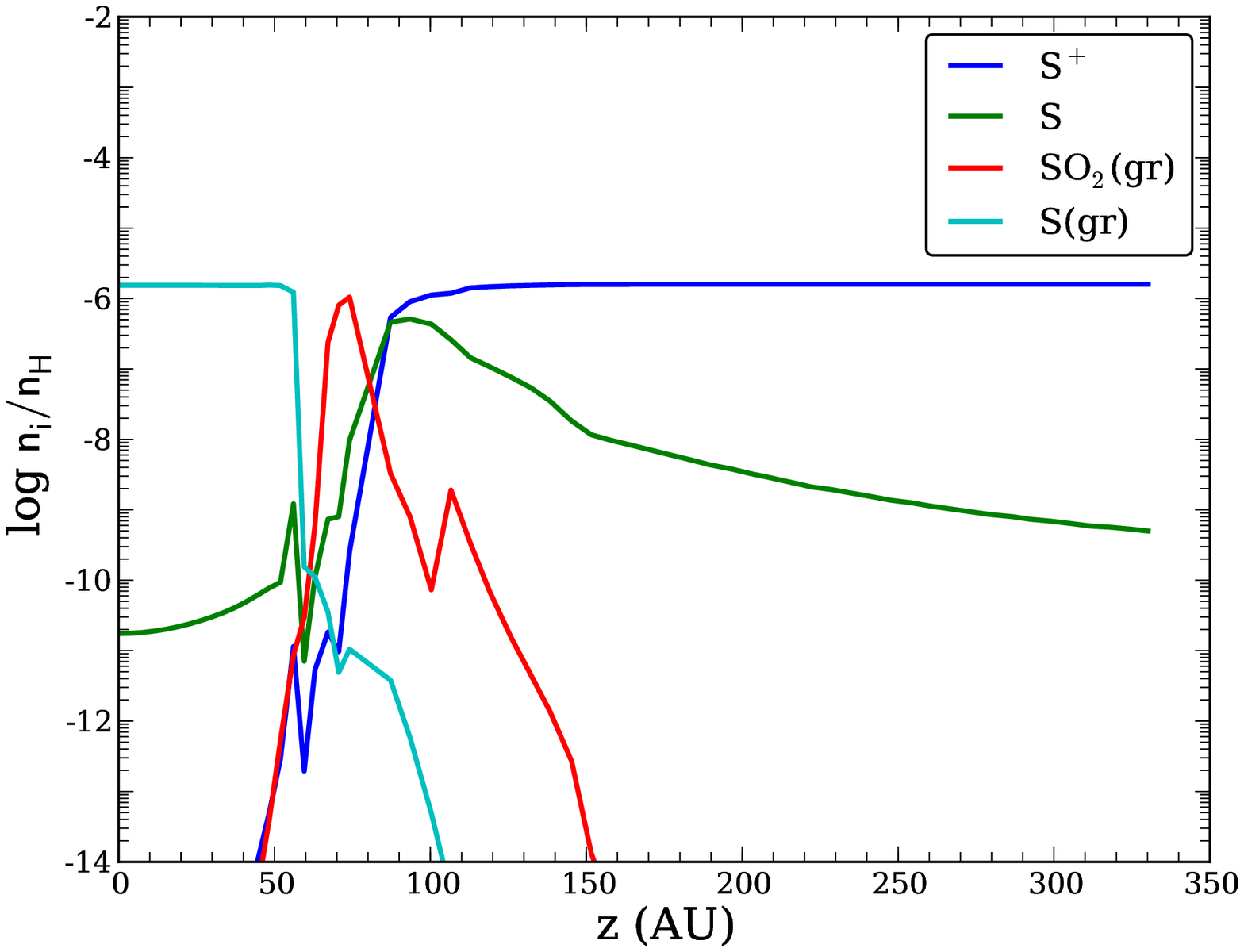}
	\caption{Plot of the vertical distribution of relative abundances of carbon-bearing (\emph{top left}), oxygen-bearing (\emph{top right}), nitrogen-bearing (\emph{bottom left}) and sulfur-bearing (\emph{bottom right}) molecules at R = 208 AU, \eps{1}.  }
	\label{line_chemistry}
\end{figure*}

The general structure of our results, demonstrated in Figure \ref{cotwoplot}, fits very well with the picture described earlier in \S \ref{section-introduction} and seen by \citet{aikawa2002}.  The figure shows C$^+$ and CO(gr) abundances for a disk with \LyA radiation and two dust settling values.  In the upper regions of the disk there is a photodissociation region (PDR) where atomic ions and neutrals exist (little CO, no CO ice), below this is a molecular layer consisting mainly of molecules (CO, no CO ice) and near the midplane is a cold region where most of the chemistry consists of molecules frozen out onto grains (no CO, CO ice).  This structure can also be seen in the upper-left plot of Figure \ref{line_chemistry}, which is a plot of the major carbon-bearing species in the disk at a radius of 208 AU vs. height.  In the PDR (z $\gtrsim$ 200 AU), all of the carbon is in the form of C$^+$ ions, in the molecular layer (50 AU $\lesssim$ z $\lesssim$ 150 AU) the carbon is in the form of CO, and in the freeze-out region (z $\lesssim$ 50 AU) the carbon is in the form of CO frozen out onto grains.  The specific location of the CO freeze-out region is sensitive to the temperature profile around 20 K, where CO freezes out onto grains.  As a result, different radiative transfer calculations will lead to slightly different locations of the freeze-out region \citep[e.g.][]{dullemond2002}.  

Similar plots can be made for the oxygen-bearing, nitrogen-bearing and sulfur-bearing chemistry in the disk (Figure \ref{line_chemistry}).  For the oxygen-bearing species, the oxygen is primarily in the form of neutral O atoms throughout the disk until the disk is cold enough for water to adsorb onto dust grains.  At that point, the oxygen quickly transitions from neutral O to H$_2$O ice on grains.  There is a small layer between these two regimes where molecules such as CO and H$_2$O exist, but the molecular layer for oxygen-bearing molecules is very narrow.  In the nitrogen-bearing chemistry, there are again three distinct regions.  The PDR consists of neutral N which transitions to N$_2$ in the molecular layer and then freezes out as N(gr) in the freeze-out region.  

The sulfur-bearing chemistry is the most complex.  While the PDR and freeze-out regions are easy to distinguish, consisting of S$^+$ and S(gr) respectively, the molecular layer is complicated by the presence of multiple sulfur-bearing species, such as SO$_2$(gr), CS(gr) and OCS(gr), that adsorb onto grains at higher temperatures than S(gr).  

\begin{figure}
	\includegraphics[width=0.5 \textwidth, clip=true, trim=10 10 0 20]{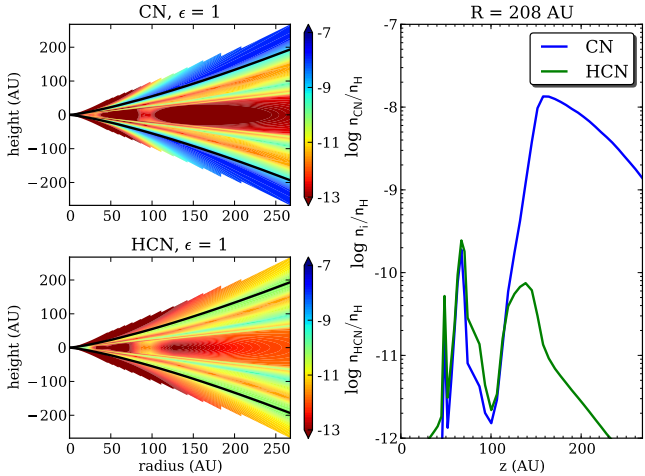}
	\caption{{\em Left Panels - } Contour plots of the abundances of CN ({\em Top}) and HCN ({\em Bottom}) for a disk model with \eps{1}.  The dashed lines indicate the $\tau$ = 1 surface in the disk.
	{\em Right Panel - } Plot of the vertical distribution of the abundance of CN and HCN in the disk at R = 208 AU, \eps{1}.  
	}
	\label{cntwoplot}
\end{figure}

Focusing on CN and HCN to demonstrate some of the more complicated chemical structure, Figure \ref{cntwoplot} shows the abundances of CN and HCN for an $\epsilon$ value of 1.  Overall we see the general structure of PDR, warm molecular layer, freeze-out; however there is also evident substructure.  The banding feature in these plots occurs due to the interaction between multiple formation and destruction mechanisms in the model.  In the band of lower abundances around z = 100 AU, the temperature is low enough that there is some freeze-out onto grains, but the UV field is still able to penetrate so that there are reactions with PDR species (C$^+$ for HCN and O for CN) as well.  Lower in the disk, the UV field is unable to penetrate as strongly, so the abundance of CN and HCN increases briefly until the temperature of the disk is cold enough for most of the CN or HCN to adsorb onto grains.  This is just one example of how the interaction between different formation and destruction reactions can complicate the chemistry of the disk.

\subsection{Dust Settling} \label{section-dustsettling}

\begin{figure}
	\includegraphics[width=0.5 \textwidth, clip=true, trim=10 10 0 20]{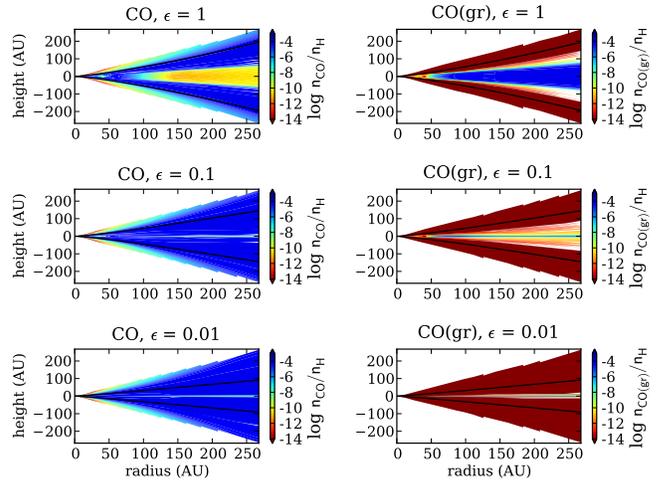}
	\caption{Contour plot of of CO and CO(gr) abundances for models with 3 different $\epsilon$ values.  The dashed lines indicate the $\tau$ = 1 surface in the disk.}
	\label{cosixplot}
\end{figure}

\begin{figure}
	\includegraphics[width=0.5 \textwidth, clip=true, trim=30 10 20 30] {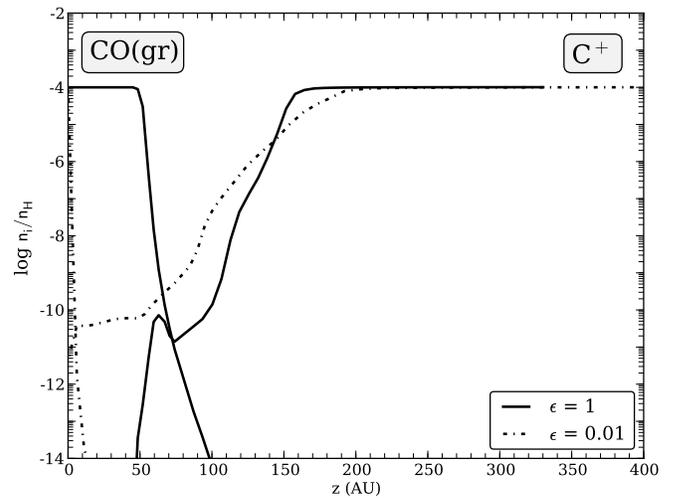}
	\caption{Plot of C$^+$ and CO(gr) vs. height at R = 208 AU for dust settling parameters of \eps{1} and \eps{0.01}.}
	\label{comultidust}
\end{figure}

An increase in dust settling will have a large impact on the thermal (Figure \ref{structure}) and chemical structure of the disk \citep{aikawa2006, jonkheid2006}.  With dust settling included most of the regions of the disk near the midplane will have temperatures higher than the 20 K needed for CO to adsorb onto dust grains \citep{hasegawa1993}.  Figure \ref{cosixplot} shows the abundance of CO and CO(gr) for $\epsilon$ values of 1, 0.1 and 0.01.  The PDR (top), molecular layer (middle) and freeze-out region (midplane) of the disk can be clearly seen in all three CO plots.  Dust settling increases the size of the molecular layer and shifts it towards the midplane at the expense of the freeze-out region.  This leads to a reduction in the overall amount of CO(gr) in the middle of the disk.  For an $\epsilon$ of 1, we see a freeze-out region in the bottom third of the disk, while for an $\epsilon$ of 0.01, only the region at the very midplane of the disk has a significant amount of CO(gr).  This effect can also be seen in Figure \ref{comultidust}, which plots C$^+$, the dominant carbon-bearing species in the PDR, and CO(gr), the dominant carbon-bearing species in the freeze-out region, for \eps{1} and 0.01 at a radius of 200 AU.  Above z $\gtrsim$ 200 AU the PDRs of the two $\epsilon$ values are similar, but the molecular layer is much larger, and the freeze-out region correspondingly smaller, in the case with dust settling included.  

\begin{figure}
	\includegraphics[width=0.5 \textwidth, clip=true, trim=60 10 10 30] {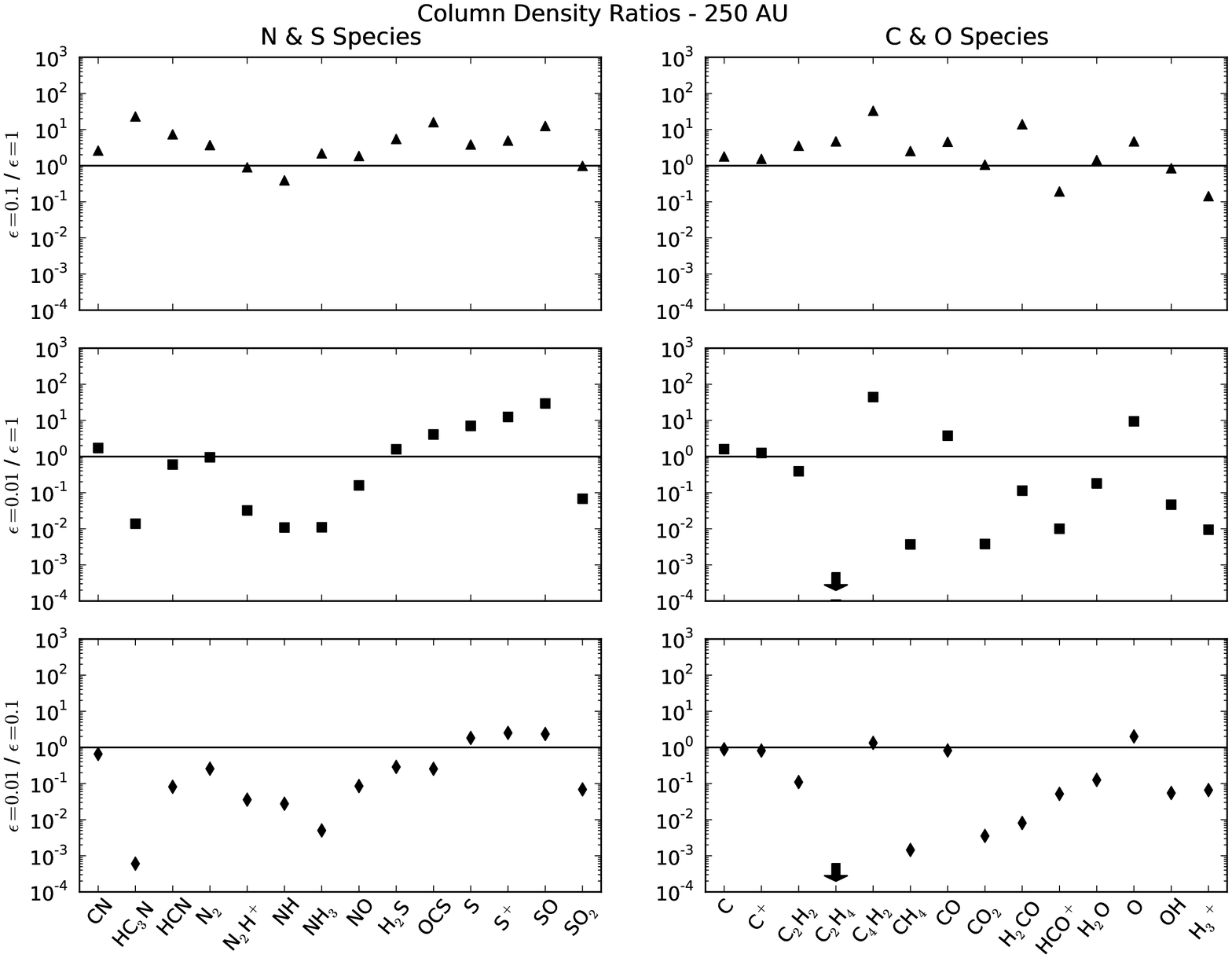}
	\caption{Column density ratios for models with different dust settling parameters, computed at 250 AU.  }
	\label{colden_dratios_250}
\end{figure}

Looking at CO allows us to illustrate the general effects of including dust settling in our model.  To quantify these effects more broadly, we will look at column density ratios for a range of representative species at the three different dust settling parameters.  Figure \ref{colden_dratios_250} compares the column densities of species at 250 AU for different dust settling parameter values.  The majority of species plotted, mostly molecular, have larger column densities for an \eps{0.1} disk than for an unsettled disk with \eps{1}.  This is especially true in the outer region of the disk and is in agreement with the results in Figure \ref{cosixplot} since the more settled disk has a larger molecular layer.  

Analyzing the results in more detail at R = 250 AU, the model with the most extreme dust settling, \eps{0.01}, shows more variations from the unsettled disk than the \eps{0.1} model does.  This is unsurprising, as more dust settling  will allow the radiation to penetrate further into the disk and affect the chemical composition more.  Looking at the nitrogen-bearing species specifically, HC$_3$N, N$_2$H$^+$, NH and NH$_3$ are all significantly depleted in the settled disk as compared with the disk with no dust settling.  In comparison, CN is slightly enhanced.  Almost all of these differences can be traced back to the increase in photodissociation rates in the settled disk.  HC$_3$N and NH$_3$ are directly photodissociated by the UV radiation, while NH is destroyed by reactions with neutral O that is produced by CO and O$_2$ photodissociation.  In comparison, N$_2$H$^+$ is destroyed through reactions with CO, which is enhanced due to the wider molecular layer in the settled disks.  Because the settled disk is warmer near the midplane, most of the nitrogen-bearing species are unable to freeze-out onto dust grains, which leads to an excess reservoir of N in the gas phase.  This also serves to increase the column density of CN, which is formed by N reacting with CH and C$_2$.  Surprisingly, HCN does not show a significant depletion in the highly settled disk, despite being strongly dissociated by UV radiation.  This is due to the formation of HCN from neutral N working to counteract the photodissociation.  

Sulfur-bearing species show a similar pattern, with the higher temperatures in the settled disk leading to very few species frozen-out onto grains.  The result of this is a significant increase in the column density of species such as S, S$^+$ and SO in the gas-phase.  SO$_2$ is strongly photodissociated, adding to the enhancement seen in SO and O.  For the carbon-bearing species, C and C$^+$ are both slightly enhanced due to the lack of frozen-out species as well as photodissociation (of CO and C respectively).  C$_2$H$_4$ and CH$_4$ are both photodissociated by the UV field as expected, leading to the significant depletion of these species in the settled disks.  

The lack of water frozen out onto grains in the settled disks due to photodesorption leads to an increase in the column density of gas-phase oxygen-bearing species such as O.  The column density of OH is not enhanced as it it destroyed due to reactions with S$^+$, which is also enhanced in the settled disk, while H$_2$O is destroyed due to photodissociation.  For some species, such as CO$_2$ and H$_2$CO, an increase in the photodissociation leads to a depletion in the column density.

\citet{aikawa2006} explored the vertical distributions of molecules in disks with larger grains and found that the peak abundances are reached at higher A$_v$ values for the cases with larger grains.  Our results, (e.g. Figure \ref{colden_dratios_250}) are in general agreement with this finding.  They found that the column densities of most molecular species did not depend on the size of the dust grains, with the exception of  HCO$^+$ and H$_3$$^+$.  The HCO$^+$ abundance was lower at higher densities, such as disks with larger grains and H$_3$$^+$ was found to be abundant in the cold midplane, which was thinner in the models with dust settling.  Our models found a similar trend for H$_3$$^+$, with the abundance decreasing as dust settling increased and the cold midplane shrank.  HCO$^+$ also showed a decrease in column density as dust settling was increased.  

\subsection{\LyA}
\begin{figure}
	\includegraphics[width=0.5 \textwidth, clip=true, trim=10 10 0 0] {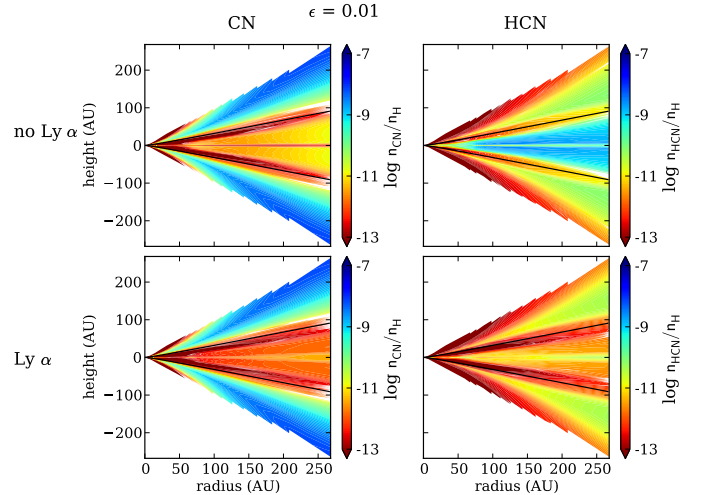}
	\caption{{\em Top Panels - } Contour plot of the abundance of CN ({\em Left}) and HCN ({\em Right}) for a model with \eps{0.01} and no \LyA radiation.  
	{\em Bottom Panels - } Contour plot of the abundance of CN ({\em Left}) and HCN ({\em Right}) for a model with \eps{0.01} and \LyA radiation included. }
	\label{cn_hcn_comp}
\end{figure}

Largely depending on whether they have a photodissociation cross section at 1216 \rm{\AA} or not, some species are dramatically affected by the presence of \LyA radiation while others are largely unaffected.  CO falls in the latter category, with no cross section at 1216 \rm{\AA} and very little change between the models run with and without \LyA radiation included.  A clear example of the interaction between \LyA radiation and a cross section at 1216 \rm{\AA} can be seen in Figure \ref{cn_hcn_comp}.  CN does not have a photodissociation cross section at 1216 \rm{\AA} and, as a result, we see very little change in abundance between the models with and without \LyA radiation included.  HCN, in comparison, has a large cross section at 1216 \rm{\AA} and is strongly depleted when \LyA radiation is included in the model.  

However, because of the numerous interconnected reactions, solely knowing the photodissociation cross section is not enough to determine the effect that adding \LyA radiation will have on the abundance of a species.  H$_2$O and OH both have photodissociation cross sections at 1216 \AA, but despite this neither species shows a significant depletion due to the presence of \LyA radiation.  Figure \ref{oh_h2o_comp} shows that OH is actually slightly enhanced and H$_2$O is slightly depleted when \LyA radiation is included in the model.  These responses are because of the importance of \LyA photodesorption.  H$_2$O(gr) will photodesorb into both H$_2$O and OH, mitigating the depletion of these species that would otherwise be seen due to photodissociation.  Because H$_2$O(gr) is the dominant oxygen-bearing species for the regions where the temperature is below 100 K, this photodesorption plays a large role in determining the predicted abundances of OH and H$_2$O.  Our treatment of photodesorption seems reasonable, as our maximum abundance for H$_2$O of a few times 10$^{-6}$ is only slightly higher than the results found in \citet{hollenbach2009} and \citet{dominik2005}.  Assuming that the H$_2$O abundance is determined solely by a balance between photodesorption and photodissociation, a reasonable assumption based on the rates of those reactions compared to the other formation and destruction reactions of H$_2$O, we can calculate an estimated abundance of H$_2$O in the disk by setting those two reaction rates equal:

\begin{figure}
	\includegraphics[width=0.5 \textwidth, clip=true, trim=10 10 0 0] {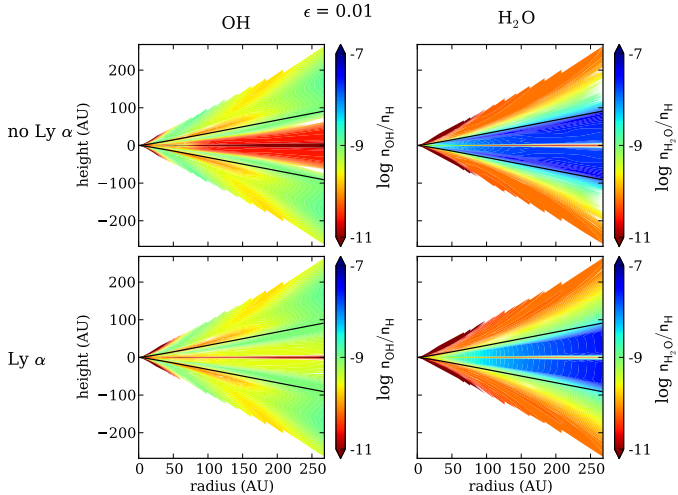}
	\caption{{\em Top Panels - } Contour plot of the abundance of OH ({\em Left}) and H$_2$O ({\em Right}) for a model with \eps{0.01} and no \LyA radiation.  
	{\em Bottom Panels - } Contour plot of the abundance of OH ({\em Left}) and H$_2$O ({\em Right}) for a model with \eps{0.01} and \LyA radiation included. }
	\label{oh_h2o_comp}
\end{figure}

\begin{multline}
n(H_2O) \times k_{photodissociation} \; (Eqn. \; (\ref{ourprate})) = \\
\shoveright{n(H_2O(gr) \times k_{photodesorption} \; (Eqn. \; (\ref{PDrate}))} \\
\shoveleft{n(H_2O) \int \sigma(\lambda) J_\lambda d\lambda =} \\
F_{UV} Y \frac{\sigma_{gr}}{N_{sites}} N_p \frac{n(H_2O(gr))}{n_{ice}} N_m^{-1} n(H_2O(gr))
\end{multline}

\noindent
Here F$_{UV}$ = $\int J_\lambda d\lambda$.  Using an average cross section ($\sigma$) for H$_2$O of $5 \times 10^{-18} cm^{-2}$, a photodesorption yield ($Y$) of $2.36 \times 10^{-3}$ and assuming that the abundance of all non H$_2$O species frozen onto grains is negligible, $n_{ice} \sim n_{H_2O(gr)}$, we calculate a maximum H$_2$O abundance of $\sim 10^{-6}$ for the case of \eps{0.1}, which matches our results.  

\begin{figure}
	\includegraphics[width=0.5 \textwidth, clip=true, trim=60 10 10 25]{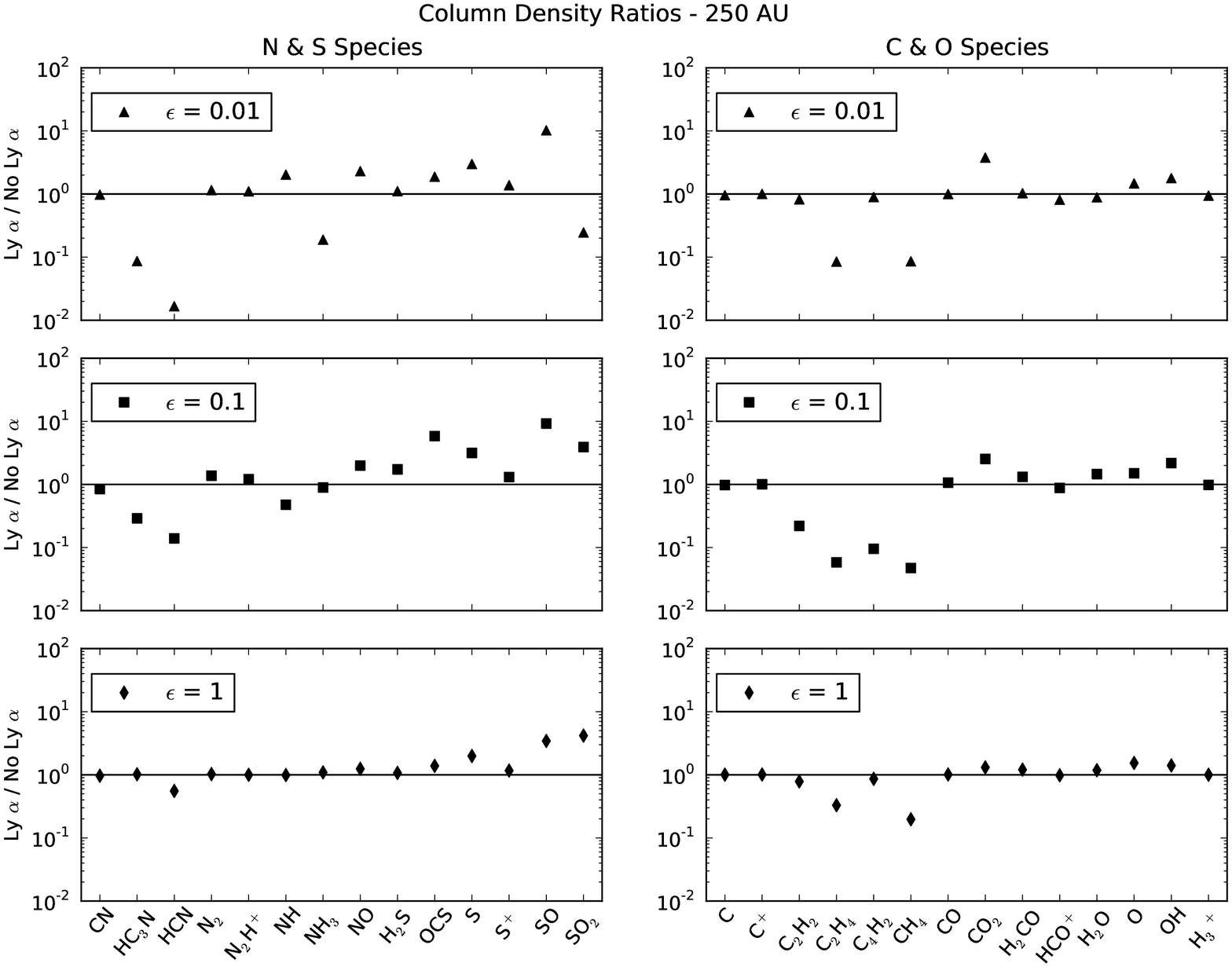}
	\caption{Column density ratios for models with and without \LyA radiation included, computed at 250 AU.}
	\label{colden_ratios_250}
\end{figure}

To more generally see the impact of \LyA radiation on the chemistry of the disk, we have plotted the ratio of column densities for models with and without \LyA radiation included for a variety of species in Figures \ref{colden_ratios_250}.  In general, the \LyA radiation has a more dramatic effect when dust settling is included in the model.  The more settled disks allow greater penetration of the \LyA radiation, thereby affecting chemistry in a much larger region of the disk.  In the case with no dust settling, since the gas and dust opacities are connected the \LyA photons are shielded before they can have much of an effect on the chemical abundances.  Unless otherwise stated, the rest of the analysis in this section will be discussing the \eps{0.01} case, as the larger effects there are easiest to see and because recent observations indicate that many protoplanetary disks have significant dust settling \citep{furlan2008}.  

Looking at the carbon-bearing species, only a couple of species are strongly depleted.  Species, such as C$_2$H$_4$ and CH$_4$, are significantly dissociated in the presence of \LyA radiation.  Other species that we would expect to be destroyed, such as C$_2$H$_2$, are not.  This lack of depletion is mostly due to an increase in the gas-phase abundance of these species that is caused by \LyA photodesorption, similar to what happens with H$_2$O.  For these species the photodesorption and photodissociation mostly balance each other out, leading to only small changes in the column densities of those molecules.  CO$_2$ is slightly enhanced due to reactions with O and OH, both of which are enhanced in the disk, which counteracts the CO$_2$ photodissociation.  

The nitrogen-bearing chemistry is similar to the carbon-bearing species.  A couple of species are photodissociated by the \LyA radiation as expected, HC$_3$N, HCN and NH$_3$, but most of the chemistry is largely unaffected.  

The sulfur-bearing species show the most dramatic differences in column densities when \LyA radiation is included.  SO$_2$, as expected, is significantly depleted in the presence of \LyA radiation due to its large photodissociation rate, though only in the highly settled disk with \eps{0.01}.  In the less settled disks, formation reactions between SO and OH or O are quick enough to counteract the photodissociation.  SO also has a large cross section at 1216 \AA, but is actually enhanced by the presence of \LyA radiation.  This occurs largely due to \LyA photodesorption of SO(gr), the photodissociation of SO$_2$ and the numerous reactions with O and S atoms, both of which are enhanced in the presence of \LyA radiation.  For OCS and S, the slight enhancement in abundances are due to \LyA photodesorption of SO which is then photodissociated to form other sulfur-bearing species.  

\begin{figure}
	\includegraphics[width=0.5 \textwidth, clip=true, trim=60 10 10 30]{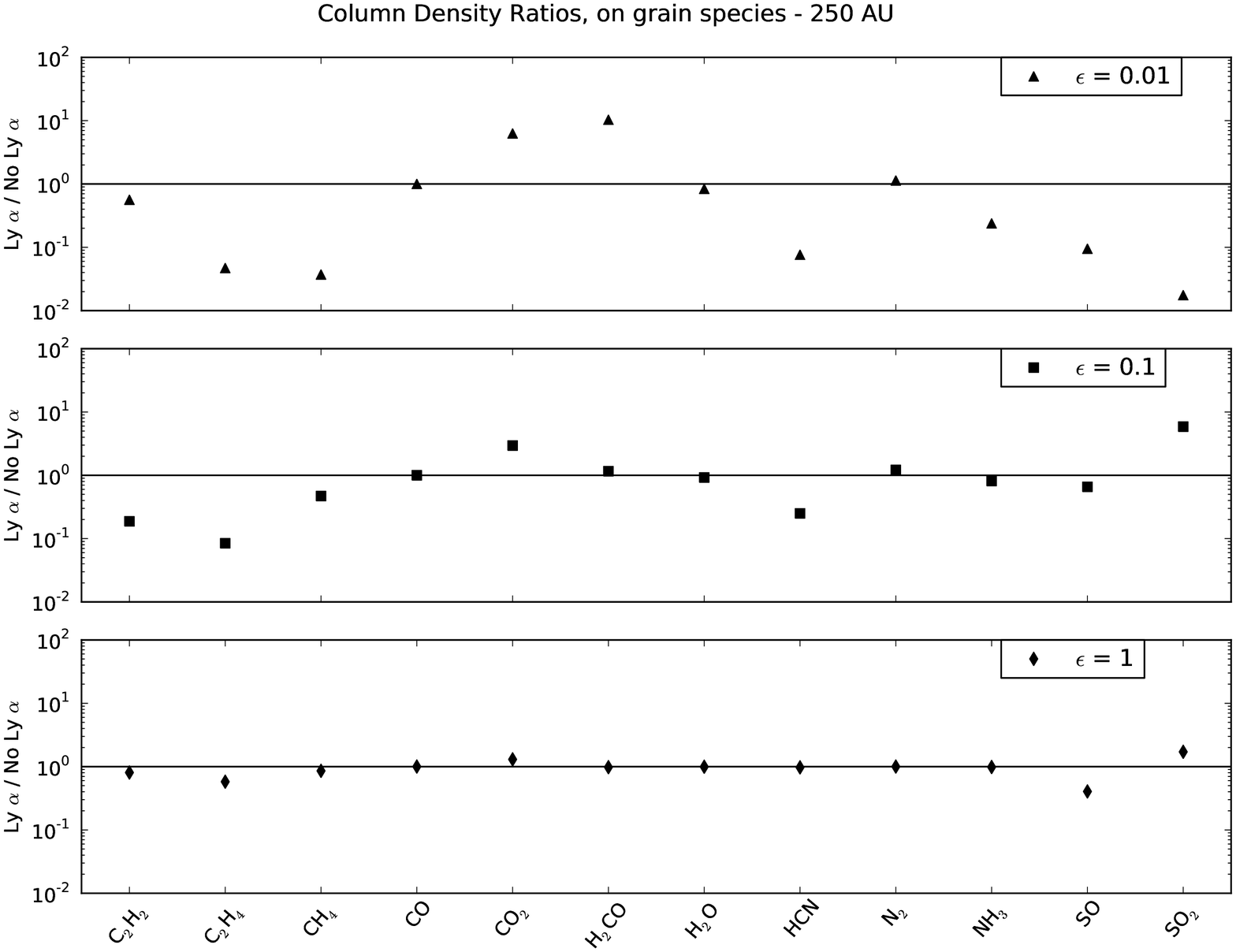}
	\caption{Column density ratios of on-grain species for models with and without \LyA radiation included, computed at 250 AU.}
	\label{colden_grratios}
\end{figure}

Due to the fact that we have very few on-grain surface reactions in our network, we expect to see a correlation between the gas-phase and on-grain abundances of species.  Specifically, species that are photodissociated in the gas-phase will be destroyed before they have a chance to freeze out onto grains.  Additionally, due to the way we dealt with photodesorption, even if those species that are sensitive to \LyA radiation did adsorb onto dust grains they would be quickly photodesorbed.  As a result, the on-grain species that are strongly affected by the presence of \LyA radiation are those that have gas phase photodissociation cross sections at 1216 \AA, with the effects being even more dramatic than for the equivalent gas-phase species.  As can be seen in Figure \ref{colden_grratios}, which plots the column density ratios of some important on-grain species at 250 AU, species such as C$_2$H$_4$, CH$_4$, HCN, NH$_3$, SO and SO$_2$ are all strongly depleted in the model with \eps{0.01}, where \LyA radiation is able to penetrate to the freeze-out region.  These species are either destroyed in the gas phase before they can freeze-out on the grains or they are photodesorbed quickly once on a grain surface.  One caveat to these results is the potential for hydrogenation of molecules on grains.  In our models we have accounted for the hydrogenation of oxygen, which is the main mechanism for water ice formation.  This water can be photodesorbed at the disk surface producing water vapor in the molecular layer.   It is possible that grain surface hydrogenation of nitrogen or carbon could perhaps lead to an increased formation of species such as NH$_3$(gr) and CH$_4$(gr), though this is not included in the current model. 

Our initial expectations were that the abundance changes that occurred would be strongly correlated with the size of a species' photodissociation cross section at 1216 \rm{\AA}.  This is clearly not the case and in fact there seems to be little to no correlation at all between the two.  The reason for this is that the interdependence of species matters more than just the cross section of a species at 1216 \rm{\AA}.  For example, the carbon chemistry in the disk is completely dominated by the reactions involving CO, as seen in Figure \ref{line_chemistry}.  In the PDR at the top of the disk, all of the carbon is in the form of C$^{+}$.  Further down in the disk where there is not enough radiation to photodissociate the CO molecule, the carbon exists in the form of CO.  Finally, at the midplane the temperatures are low enough for CO to freeze out and the carbon is in the form of CO ice on the grains.  There is an interesting region just between the molecular and freeze-out regions in which the temperature is too high for CO to freeze out, but cold enough for other molecules to adsorb onto dust grains.  In this region we start to see the presence of more complex molecules frozen out onto the grains, such as C3, CH$_4$, HCN and C$_2$H$_2$.  Even though the reactions that form these molecules are fairly slow compared with the rest of the carbon chemistry, since they are being depleted from the gas phase through adsorption there is enough time to build up a substantial abundance of these more complex species and remove that carbon from the rest of the network.  

CO does not have a photodissociation cross section at 1216 \rm{\AA}, which means that any species connected to CO will have only minor changes in abundances when \LyA radiation is included.  This effect can be seen clearly at 250 AU for species such as HCO$^+$, which is closely connected to CO, and CH$_{4}$, which is more weakly connected.  The HCO$^+$ abundance is largely unaffected by the presence of \LyA radiation, while the CH$_{4}$ is strongly depleted.  

The oxygen-bearing chemistry in the disk is actually fairly simple in that, like the carbon-bearing chemistry, it is dominated by a single species, H$_2$O.  As seen in Figure \ref{line_chemistry}, the PDR at the top of the disk is made up completely of free O atoms, which freeze out as water ice as soon as the disk is cold enough to allow it, around 100 K.  There is a little oxygen in CO and CO(gr) in the molecular and freeze-out regions respectively, but these are less abundant by a factor of a few than the O and H$_{2}$O(gr).  

The nitrogen-bearing chemistry is similarly straight-forward.  All of the nitrogen in our model starts in the form of N$_2$, but is quickly dissociated to form neutral N in the PDR.  Neither N$_2$ nor N has a large photodissociation cross section at 1216 \rm{\AA}, so this process is largely unaffected by the presence of \LyA radiation in the UV field.  In the molecular layer, a range of nitrogen-bearing species exist, including NH$_3$ and HCN.  Both of these have strong photodissociation cross sections at 1216 \rm{\AA}, so they are destroyed in the highly settled disk models leading instead to a molecular layer made up primarily of N$_{2}$.  Near the midplane, the nitrogen exists in the form of N(gr) when the temperature is low enough for it to freeze out, which can be seen in Figure \ref{line_chemistry}.  

The sulfur chemistry is a little more complicated.  Figure \ref{line_chemistry} indicates that  the primary sulfur-bearing species are S$^+$ in the PDR and S(gr) in the midplane.  However, the sulfur species with large cross-sections at 1216 \AA\ are SO and SO$_2$.  As was discussed above, the photodissociation of SO$_{2}$ leads to the formation of SO, which is also formed through photodesorption and reactions between O and S atoms.  This means that despite the large cross section at 1216 \AA, SO is actually enhanced in the presence of \LyA radiation, though the SO(gr) abundance is depleted.  

\subsection{Comparison with Observations}

\begin{figure*}
	\includegraphics[width=0.5 \textwidth, clip=true, trim=20 5 40 20]{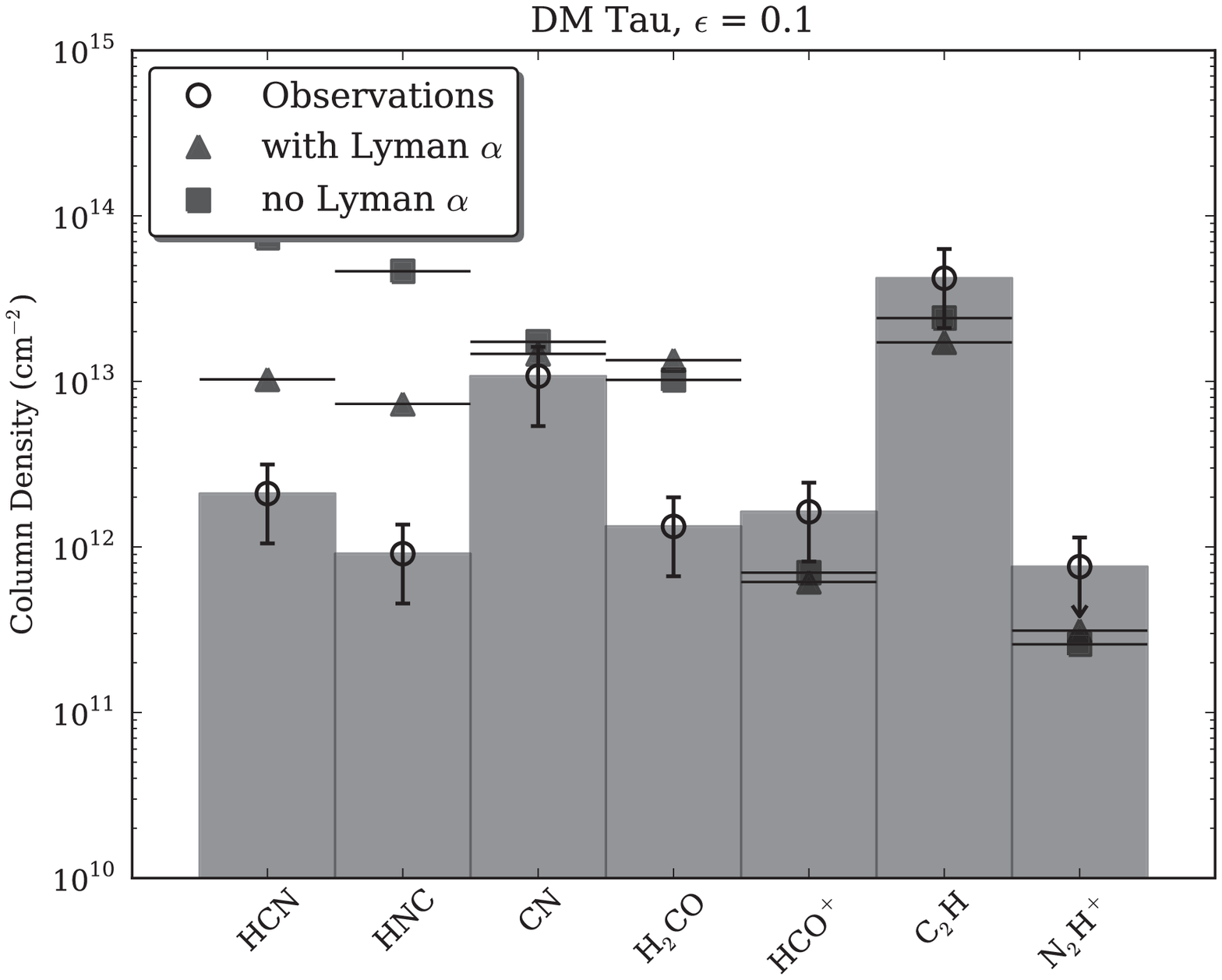}
	\includegraphics[width=0.5 \textwidth, clip=true, trim=20 5 40 20]{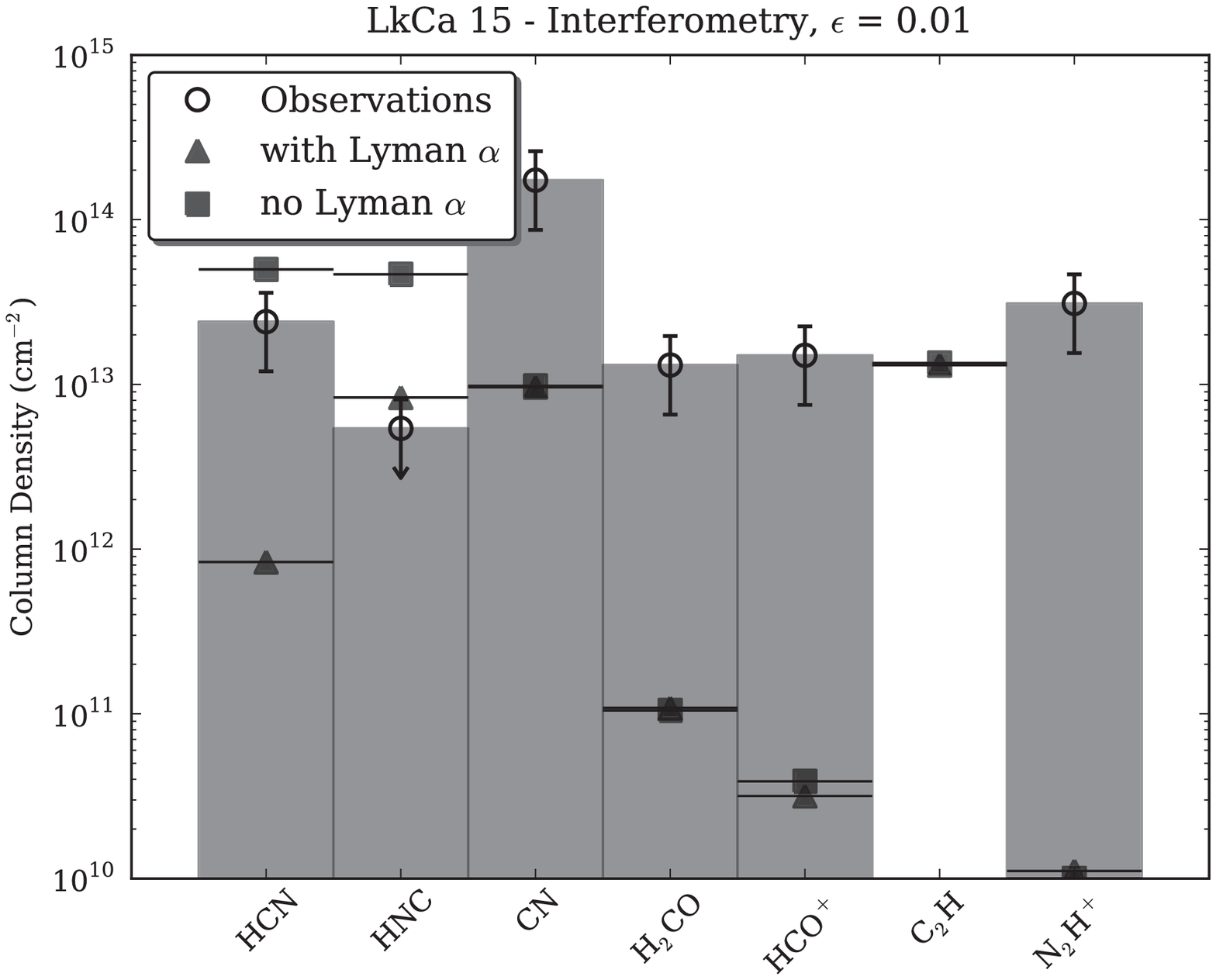}
	\includegraphics[width=0.5 \textwidth, clip=true, trim=20 10 40 10]{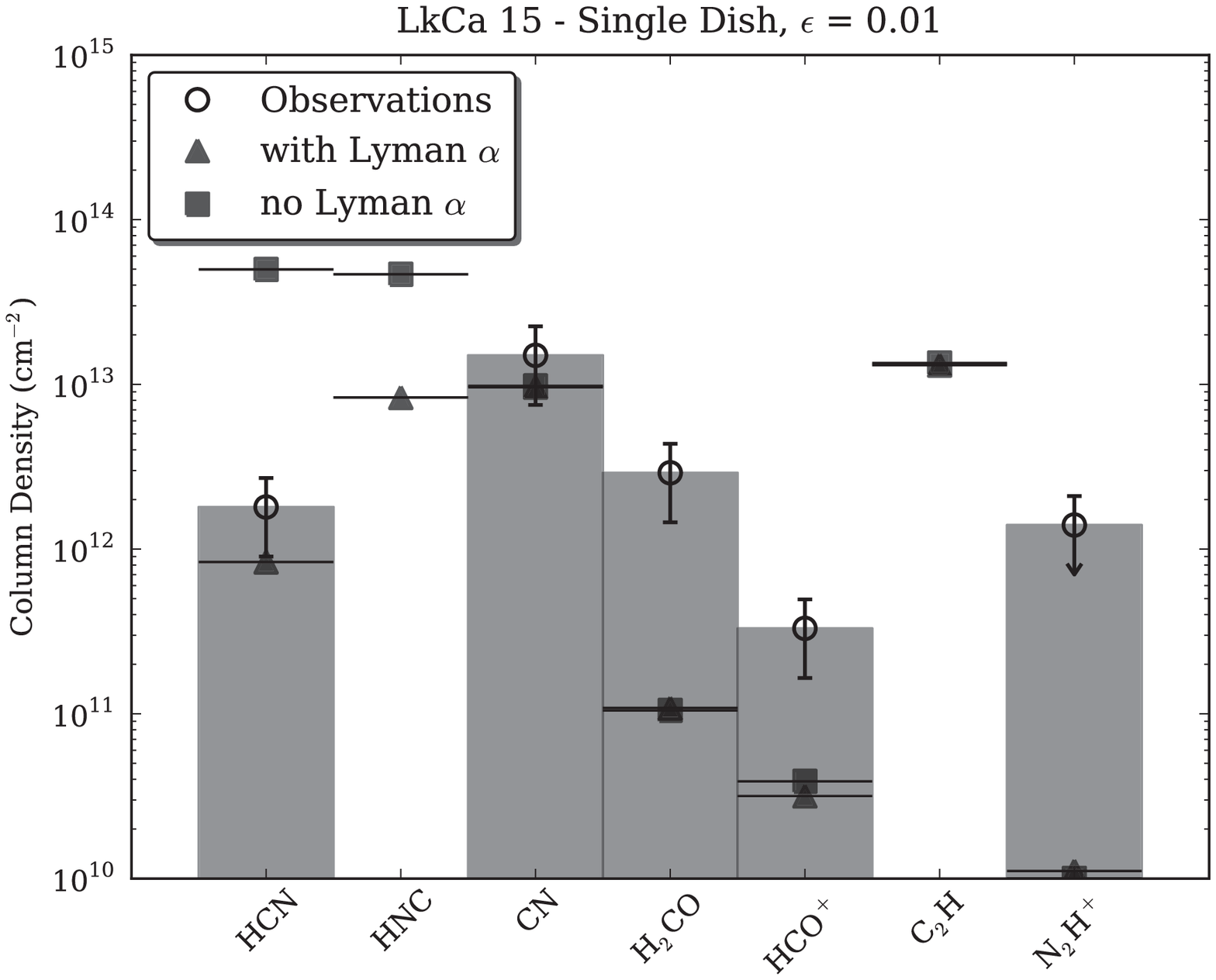}
	\includegraphics[width=0.5 \textwidth, clip=true, trim=20 10 40 10]{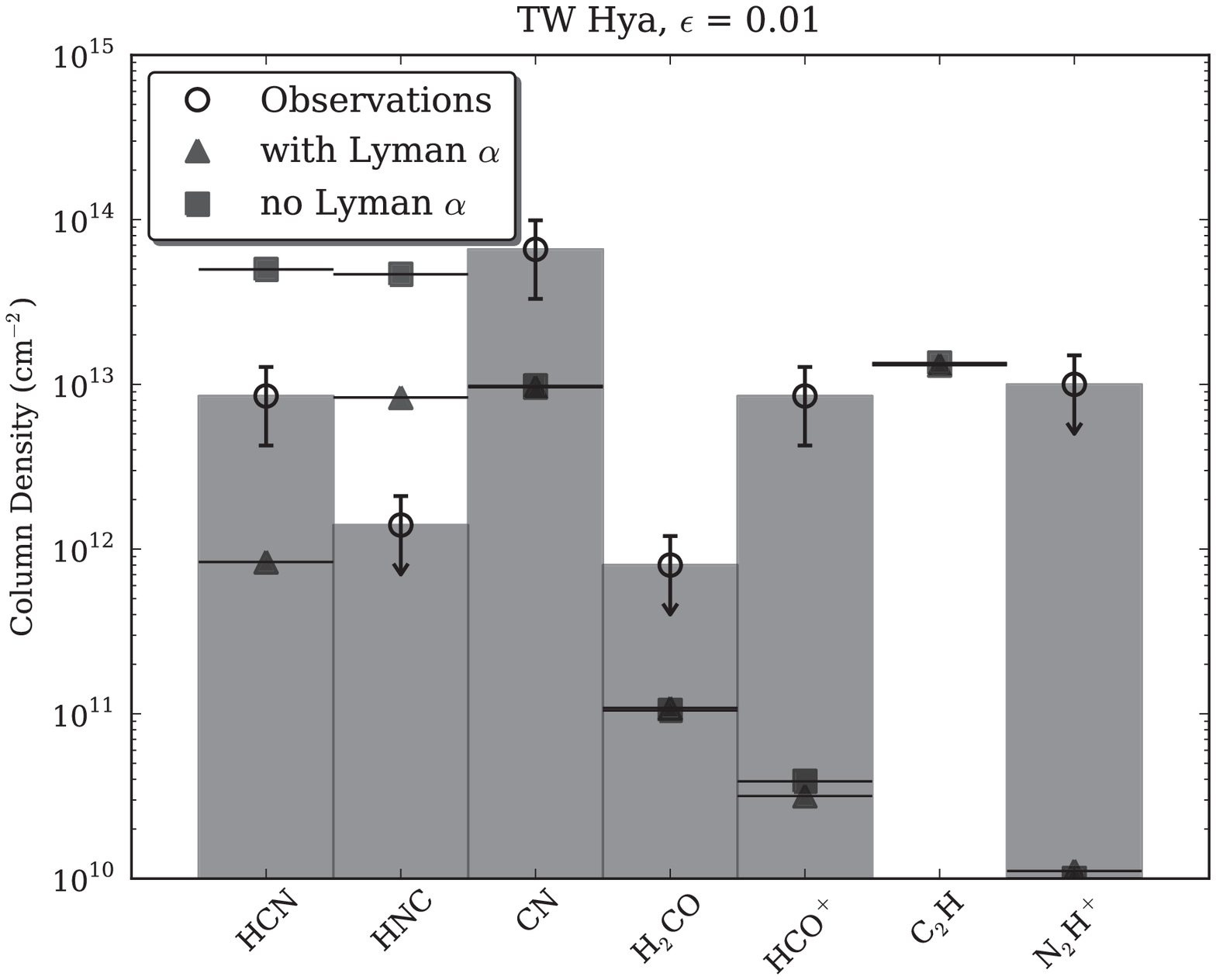}
	\caption{Figures comparing our model with \eps{0.01} (except for DM Tau, which is \eps{0.1}) and observations of three disks .  The gray bars and open circles represent the observations, with error bars of 50\%.  The triangles show our model with \LyA radiation included and the squares show our model with \LyA radiation not included.  
	\newline \noindent \emph{top left} - DM Tau \citep{aikawa2002}, \eps{0.1}.
	\newline \noindent \emph{top right} - LkCa 15, Interferometric observations \citep{qi2001, qi2003, aikawa2003}.
	\newline \noindent \emph{bottom left} - LkCa 15, single-dish observations \citep{thi2004}.
	\newline \noindent \emph{bottom right} - TW Hya \citep{thi2004, van-dishoeck2003}.}
	\label{obscomp}
\end{figure*}

We have primarily focused on the outer parts of the protoplanetary disk in this paper because that is the region of the disk probed by current millimeter and sub-millimeter observations.  In Table \ref{tbl-obscomp} we compare the results of our model with observations taken of three disks: DM Tau, LkCa 15 and TW Hya.  The column densities listed for our model were taken at a radius of 250 AU and a time of 10$^6$ years for each of the dust settling parameters.  For each species in the table, the top line includes \LyA radiation while the second line does not.  In addition, we present figures of the best fit models vs. the observations with and without \LyA radiation.  These can be seen in Figure \ref{obscomp}.  

\begin{deluxetable*}{llllrrrr}
\tablewidth{0pt}
\tabletypesize{\small}
\tablecaption{Comparison of the Model with Observations \label{tbl-obscomp}}
\tablehead{
 & \multicolumn{3}{c}{Models} &  & \multicolumn{2}{c}{LkCa 15} \\
\colhead{Species} 
&\colhead{\eps{1}} & \colhead{\eps{0.1}} & \colhead{\eps{0.01}} 
& \colhead{DM Tau\tablenotemark{a}} & \colhead{Interferometer\tablenotemark{b}} & \colhead{Single dish\tablenotemark{c}} & \colhead{TW Hya\tablenotemark{c}}
}

\startdata
H$_2$		& 8.3 (22) & 3.8 (22) & 3.1 (22) &
	3.8 (21) & \nodata & \nodata & \nodata \\
			& 8.3 (22) & 3.8 (22) & 3.1 (22) \\ 

CO			& 1.3 (18) & 6.2 (18) & 5.1 (18) &
	5.7 (16)  &  1.7 (18)\tablenotemark{d}  &  1.9 (16)  &  3.2 (16)  \\
			& 1.3 (18) & 5.8 (18) & 5.1 (18) \\ 

HCN			& 1.4 (12) & 1.0 (13) & 8.3 (11) &
	2.1 (12)  &  2.4 (13)  &  1.8 (12)  &  8.5 (12)\tablenotemark{e} \\
			& 2.5 (12) & 7.4 (13) & 5.0 (13) \\ 
			
HNC			& 1.4 (12) & 7.3 (12) & 8.3 (12) &
	9.1 (11)  &  $<$5.4 (12)  &  \nodata  &  $<$1.4 (12) \\
			& 2.6 (12) & 4.6 (13) & 4.7 (13) \\ 

CN			& 5.6 (12) & 1.5 (13) & 9.6 (12) &
	9.5-12 (12)  &  9.7-25 (13)  &  1.5 (13)  &  6.6 (13) \\
			& 5.8 (12) & 1.7 (13) & 9.8 (12) \\ 

H$_2$CO		& 9.6 (11) & 1.3 (13) & 1.1 (11) &
	7.6-19 (11)  &  7.2-19 (12)\tablenotemark{f}  &  7.1-51 (11)  &  $<$8.0 (11) \\
			& 7.9 (11) & 1.0 (13) & 1.0 (11) \\ 

HCO$^+$		& 3.2 (12) & 6.1 (11) & 3.2 (10) &
	4.6-28 (11)  &  1.5 (13)  &  3.3 (11)  &  8.5 (12)\tablenotemark{e} \\
			& 3.2 (12) & 7.0 (11) & 3.9 (10) \\ 

C$_2$H		& 6.6 (12) & 1.7 (13) & 1.3 (13) &
	4.2 (13)  &  \nodata  &  \nodata  &  \nodata \\
			& 6.9 (12) & 2.4 (13) & 1.3 (13) \\ 

N$_2$H$^+$	& 3.5 (11) & 3.1 (11) & 1.1 (10) &
	$<$7.6 (11)  &  3.1 (13)\tablenotemark{d}  &  $<$1.4 (12)  &  $<$1.0 (13) \\
			& 3.5 (11) & 2.6 (11) & 1.0 (10) \\ 

\enddata

\tablecomments{Model column densities (cm$^{-2}$) are calculated at t = $\rm{1 \times 10^6}$ yr, R = 250 AU.  For each species, the first row includes \LyA radiation while the second line does not.  The table of observations was compiled by \citet{willacy2007}.}
\tablenotetext{a}{\citet[][derived from \citealt{dutrey1997}]{aikawa2002}}
\tablenotetext{b}{\citet{qi2001} unless otherwise noted}
\tablenotetext{c}{\citet{thi2004} unless otherwise noted}
\tablenotetext{d}{\citet{qi2003}}
\tablenotetext{e}{\citet{van-dishoeck2003}}
\tablenotetext{f}{\citet{aikawa2003}}

\end{deluxetable*}

We did not produce source-specific models for each of these objects, but instead show our results from the generic T Tauri star model described in \S \ref{physical-model}.  The objects listed here are all transition disks, which generally have lower accretion rates and larger disk masses \citep{najita2007}.  Despite this, we feel that our models can still provide a decent approximation to the observations for two reasons.  First, $\dot{M} / \alpha$, which determines the density structure, is roughly consistent between our model and these objects.  For our model, $\dot{M} = 1.0 \times 10^{-8}$ and $\alpha = 1.0 \times 10^{-2}$ so $\dot{M} / \alpha = 1.0 \times 10^{-6}$.  Second, the column densities derived from observations are very sensitive to assumptions about the physical structure, adding a level of uncertainty to those values.  For most of the species shown here the addition of \LyA radiation does not affect the column densities much.  However, for the species that are affected by \LyA radiation, HCN and HNC, the models with \LyA radiation included often produce a better agreement with the observations than those without it.  The best fit model to each set of observations was determined using a two-sample ks-test of the observations with each of the 6 models (3 dust values, with and without \LyA radiation).  In all cases the statistics are severely limited as the best fit is determined by so few data points.  

{\em DM Tau - } \citet{calvet2005} determined that this object has a stellar mass of 0.65 M$_\Sun$, a stellar radius of 1.2 R$_\Sun$, a stellar temperature of 3720 K, an accretion rate of 2 $\times$ 10$^{-9}$ M$_\Sun$ yr$^{-1}$ and an $\alpha$ of 0.0015, leading to an $\dot{M} / \alpha$ of 1.33 $\times$ 10$^{-6}$.  It has a disk mass of 0.05 M$_\Sun$ and the SED was best fit with \eps{0.1}.  The observations are from \citet{dutrey1997} who determined the gas density distribution of the disk and used this to derive the average fractional abundances with respect to H2, assuming that the fractional abundances were the same everywhere in the disk.  \citet{aikawa2002} used this data and integrated vertically to get the observed column densities listed in Table \ref{tbl-obscomp}.  Our models for \eps{0.1} produced the best fit to the data, with the model without \LyA radiation included actually fitting the data slightly better, though the ks-test for the models with and without \LyA radiation were similar (Figure \ref{obscomp}, top left).

{\em LkCa 15 - } \citet{espaillat2007} found that this object has a stellar mass of 1.1 M$_\Sun$, a stellar radius of 1.7 R$_\Sun$, a stellar temperature of 4350 K, an accretion rate of 2.4 $\times$ 10$^{-9}$ M$_\Sun$ yr$^{-1}$ and an $\alpha$ of 0.0006, leading to an $\dot{M} / \alpha$ of 4.0 $\times$ 10$^{-6}$.  It has a disk mass of 0.1 M$_\Sun$ and the best fit SED model was with \eps{0.001}.  Looking at our results for \eps{0.01} we find that our models are actually a slightly poorer fit to the interferometric data when \LyA is included (Figure \ref{obscomp}, top right), but our models with \LyA radiation included and \eps{0.01} produced the best fit to the single dish observations (Figure \ref{obscomp}, bottom left).  One of the largest differences between our model and the single-dish data occurs in the column densities of H$_2$CO, where our models significantly underestimate the observed values.  This is not unexpected, as \citet{oberg2009-CH3OH} has suggested that on-grain reactions can be a critical formation pathway for H$_2$CO, which is not currently included in our model.  

{\em TW Hya - } This object has been studied extensively and has been found to have a stellar mass of 0.6 M$_\Sun$, a stellar radius of 1.0 R$_\Sun$, a stellar temperature of 4000 K \citep{calvet2002}, an accretion rate of 3.5 $\times$10$^{-9}$ M$_\Sun$ yr$^{-1}$ \citep{espaillat-privcom} and an $\alpha$ of 0.0025 \citep{espaillat-privcom}, leading to an $\dot{M} / \alpha$ of 1.4 $\times$ 10$^{-6}$.  It has a disk mass of 0.06 M$_\Sun$ \citep{calvet2002} and the best fit SED model was with \eps{0.01} \citep{espaillat-privcom}.  For the observed column densities, the disk was assumed to be 165 AU in radius.  Our model with \eps{0.01} and no \LyA radiation included produced the best fit to the observations, though again the fit for the model with and without \LyA radiation were very similar (Figure \ref{obscomp}, bottom right).  In this case the inclusion of \LyA radiation actually produces too little HCN in the model.  

Column densities for a larger subset of species from our model can be seen in Table \ref{tbl-colden} for \eps{0.01}, R = 250 AU and time steps of $3 \times 10^{5}$ and $1 \times 10^{6}$ years.

\begin{deluxetable}{lll}
\tablewidth{0pt}
\tabletypesize{\small}
\tablecaption{Column Densities at \eps{0.01}, R = 250 AU\tablenotemark{a} \label{tbl-colden}}
\tablehead{
\colhead{Species} & \colhead{$3 \times 10^{5}$ yrs} & \colhead{$1 \times 10^{6}$ yrs}
}
\startdata
C & $1.396 \times 10^{15}$ & $1.399 \times 10^{15}$ \\ 
C$^+$ & $3.529 \times 10^{16}$ & $3.546 \times 10^{16}$ \\ 
C$_2$H$_2$ & $2.143 \times 10^{11}$ & $2.144 \times 10^{11}$ \\ 
C$_2$H$_4$ & $1.987 \times 10^{4}$ & $1.932 \times 10^{4}$ \\ 
C$_3$H$_2$ & $4.768 \times 10^{10}$ & $4.776 \times 10^{10}$ \\ 
C$_4$H$_2$ & $9.077 \times 10^{9}$ & $9.098 \times 10^{9}$ \\ 
CH$_4$ & $1.005 \times 10^{11}$ & $9.976 \times 10^{10}$ \\ 
CN & $9.621 \times 10^{12}$ & $9.613 \times 10^{12}$ \\ 
CO & $5.083 \times 10^{18}$ & $5.082 \times 10^{18}$ \\ 
CO$_2$ & $4.471 \times 10^{10}$ & $4.715 \times 10^{10}$ \\ 
H$_2$ & $3.126 \times 10^{22}$ & $3.126 \times 10^{22}$ \\ 
H$_2$CO & $1.077 \times 10^{11}$ & $1.084 \times 10^{11}$ \\ 
H$_2$O & $8.086 \times 10^{14}$ & $8.095 \times 10^{14}$ \\ 
H$_2$S & $1.191 \times 10^{11}$ & $1.154 \times 10^{11}$ \\ 
H$_3$$^+$ & $1.540 \times 10^{11}$ & $1.590 \times 10^{11}$ \\ 
HC$_3$N & $1.289 \times 10^{7}$ & $1.280 \times 10^{7}$ \\ 
HCN & $1.060 \times 10^{12}$ & $8.341 \times 10^{11}$ \\ 
HCO$^+$ & $3.086 \times 10^{10}$ & $3.170 \times 10^{10}$ \\ 
N & $9.914 \times 10^{17}$ & $1.020 \times 10^{18}$ \\ 
N$_2$ & $1.529 \times 10^{17}$ & $1.374 \times 10^{17}$ \\ 
N$_2$H$^+$ & $1.153 \times 10^{10}$ & $1.112 \times 10^{10}$ \\ 
NH & $1.055 \times 10^{11}$ & $7.511 \times 10^{10}$ \\ 
NH$_3$ & $4.581 \times 10^{10}$ & $3.611 \times 10^{10}$ \\ 
NO & $2.575 \times 10^{13}$ & $2.567 \times 10^{13}$ \\ 
O & $5.688 \times 10^{18}$ & $5.778 \times 10^{18}$ \\ 
O$_2$ & $5.358 \times 10^{14}$ & $5.707 \times 10^{14}$ \\ 
OCS & $1.717 \times 10^{11}$ & $1.516 \times 10^{11}$ \\ 
OH & $2.736 \times 10^{13}$ & $2.908 \times 10^{13}$ \\ 
S & $1.067 \times 10^{16}$ & $8.180 \times 10^{15}$ \\ 
S$^+$ & $6.230 \times 10^{16}$ & $5.676 \times 10^{16}$ \\ 
S$_2$ & $3.623 \times 10^{12}$ & $1.345 \times 10^{12}$ \\ 
SO & $1.126 \times 10^{15}$ & $9.567 \times 10^{14}$ \\ 
SO$_2$ & $8.016 \times 10^{11}$ & $7.548 \times 10^{11}$ \\ 
\enddata
\tablenotetext{a}{Column Densities are all in cm$^{-2}$}
\end{deluxetable}

\section{Discussion \& Summary} \label{section-discussion}

The existing datasets are quite sparse and our ability to compare between model and observations is somewhat hampered by the methods chosen for column density determination.   With these caveats in mind, within our models the best match is obtained with both dust settling and \LyA radiation included.  Since both of these features are known to be present in accreting gas-rich disk systems, in particular TW Hya, DM Tau, GM Aur, and LkCa15, we believe that these physical effects need to be incorporated into protoplanetary disk chemistry models.  The dust settling had the expected effect of increasing the UV penetration as the dust settling increases \citep{dullemond2004}.  This changed the size and location of the molecular layer and freeze-out regions, which in turn affected the column densities of molecules in the disk.  Most of the differences that occurred between the different dust settling models came about due to an increase in photodissociation reaction rates as dust settling increased.  In addition, the smaller freeze-out regions of the settled disks allowed species such as N$_2$H$^+$ to be depleted due to the increased molecular layers in those disks.  Some of these changes were multiple orders of magnitude effects and since most disks appear to be heavily settled \citep{dalessio2006, furlan2006} leaving dust settling out of protoplanetary disk models will produce results that are less likely to match observations.  For the few objects where we were able to compare our results with observations, all three disks were fit best with a disk where some dust settling was included.  

As with dust settling, the inclusion of \LyA radiation transfer to the protoplanetary disk model had a significant effect on the resulting chemical abundances.  While not all species with large photodissociation cross sections at 1216 \rm{\AA} were heavily depleted, many species had column densities that were changed by an order of magnitude or more.  Some of the most depleted species included HC$_3$N, HCN, NH$_3$, SO$_2$, C$_2$H$_4$ and CH$_4$, all of which were strongly photodissociated by the inclusion of \LyA radiation.  In addition, some species had column densities that were enhanced, such as S, SO and CO$_2$, due to the photodissociation of other species with cross sections at 1216 \AA.

For some species, the effects of \LyA induced photodesorption played a large role in determining how that species was affected by the inclusion of \LyA radiation.  Species such as H$_2$O and OH had their abundances increased due to \LyA photodesorption which counteracted the destruction of these molecules by photodissociation. 

Comparing the results from our models with the observations of three disk systems indicated that in general the models with \LyA radiation included produced similar or slightly better fits to the data.  However the statistics on these fits were very limited as only a handful of species were observed and only two of those were species that were affected by the presence of \LyA radiation.  Despite this, we feel it is promising that the models with \LyA radiation included in them often produced better fits than those without \LyA radiation.  

As mentioned previously, the most molecule rich disks all have indications of either strong \LyA emission (e.g. TW Hya) or broad \LyA wing emission (e.g. DM Tau, GM Aur, LkCa 15).  In both of these cases there is also indirect evidence for \LyA emission through fluorescent lines of molecular hydrogen.  Therefore we believe that \LyA emission is an intrinsic feature of most, if not all, accreting disks.  Our models have shown that the inclusion of both \LyA radiation (and specifically \LyA radiation transfer) and dust settling are important components for understanding disk chemistry.  At present the observational data do not allow for definitive statements regarding which physical effect dominates, but this will change as we head into the age of ALMA.  

\section{Acknowledgments}
We would like to thank Paola D'Alessio for the generous use of her protoplanetary disk models.  This work was supported by the National Science Foundation under Grant 0707777.  Dmitry Semenov acknowledges support by the {\it Deutsche Forschungsgemeinschaft} through SPP~1385: "The first ten million years of the solar system - a planetary materials approach'' (SE 1962/1-1)."

\bibliographystyle{apj}
\bibliography{jeffrey-bibdesk}

\end{document}